\documentclass[11pt,a4paper]{article}
\pdfoutput=1
\usepackage{cmap}
\usepackage[T1]{fontenc}
\usepackage[utf8]{inputenc}
\usepackage{mathbbol}
\usepackage{booktabs}
\usepackage{bbm}
\usepackage{t1enc}
\usepackage{subfigure}
\usepackage{dsfont}
\usepackage{slashed}
\usepackage{multirow}
\usepackage{lscape}
\usepackage{wrapfig}
\usepackage{amsmath}
\usepackage{amssymb}
\usepackage{mathtools}

\usepackage{setspace}
\usepackage{lscape}
\usepackage{rotfloat}
\usepackage{rotating}
\usepackage{caption}
\usepackage{longtable}
\usepackage{jheppub}
\usepackage{cleveref}
\newcommand{\be}{\begin{equation}}
\newcommand{\ee}{\end{equation}}
\newcommand{\Tr}{\textmd{Tr}}
\newcommand{\sym}{\textmd{sym}}

\newcommand{\dMsq}{\delta{\mathbb{M}}^2}
\newcommand{\Mbarsq}{\bar{\mathbb{M}}^2}
\newcommand{\MbarH}{\bar{\mathbb{M}}_{\rm H}}

\long\def\symbolfootnote[#1]#2{\begingroup%
\def\thefootnote{\fnsymbol{footnote}}\footnote[#1]{#2}\endgroup} 

\hypersetup{%
	pdftitle    = {D and Ds decay constants in Nf=2+1 QCD with Wilson fermions},
	pdfauthor   = {S. Kuberski, F. Joswig, S. Collins, J. Heitger, W. Söldner},
}%

\title{\boldmath $\mathrm{D}$ and $\mathrm{D_s}$ decay constants in $N_{\rm f}=2+1$ QCD with Wilson fermions}

\collaborationImg{
	\includegraphics[height=2.2em]{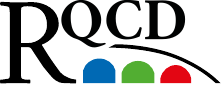} Collaboration \hspace{1em}
	\includegraphics[height=2.2em]{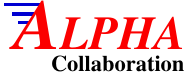}}
\author[a,b,c]{Simon~Kuberski,}
\author[d]{Fabian~Joswig,}
\author[e]{Sara~Collins,}
\author[f]{Jochen~Heitger,}
\author[e]{Wolfgang~S\"oldner}

\affiliation[a]{Theoretical Physics Department, CERN, 1211 Geneva 23, Switzerland.}
\affiliation[b]{Helmholtz-Institut Mainz, Johannes Gutenberg-Universität Mainz, Staudingerweg 18, \\55128 Mainz, Germany.}
\affiliation[c]{GSI Helmholtzzentrum für Schwerionenforschung, Planckstraße 1, 64291 Darmstadt, Germany.}
\affiliation[d]{Higgs Centre for Theoretical Physics, School of Physics and Astronomy, \\The University of Edinburgh, Edinburgh EH9 3FD, UK.}
\affiliation[e]{Institut f\"ur Theoretische Physik, Universit\"at Regensburg, 93040 Regensburg, Germany.}
\affiliation[f]{Institut f\"ur Theoretische Physik, Universit\"at M\"unster, Wilhelm-Klemm-Stra{\ss}e 9, 48149 M\"unster, Germany.}

\emailAdd{simon.kuberski@cern.ch}
\emailAdd{fabian.joswig@ed.ac.uk}
\emailAdd{sara.collins@ur.de}
\emailAdd{heitger@uni-muenster.de}
\emailAdd{wolfgang.soeldner@ur.de}

\abstract{We present results for the leptonic decay constants of the D and D$_{\rm s}$ mesons from $N_{\rm f}=2+1$ lattice QCD. We employ a set of 49 high statistics gauge ensembles generated by the Coordinated Lattice Simulations~(CLS) effort utilising non-perturbatively improved Wilson fermions and the tree-level Symanzik improved gauge action at six values of the lattice spacing in the range $a = 0.098$\,fm down to $a = 0.039$\,fm, with pion masses varying from around $420\,$MeV down to below the physical point. The ensembles lie on three trajectories in the quark mass plane, two trajectories intersecting close to the physical quark mass point and the third one approaching the SU(3) chiral limit, enabling tight control of the light and strange quark mass dependence. We obtain $f_{\mathrm{D_s}}=246.8(1.3)\,$MeV, $f_\mathrm{D}=208.4(1.5)\,$MeV and $f_{\mathrm{D_s}}/f_\mathrm{D}=1.1842(36)$, where the precision of our results is mostly limited by the determination of the scale.
}


\dedicated{\normalfont CERN-TH-2024-052, MITP-24-047, MS-TP-24-09}

\begin{document}

\maketitle

\section{Introduction \label{s:intro}}

Flavour physics plays an important role in searches for signals of
beyond the Standard Model interactions. In particular, non-unitarity of the
Cabibbo-Kobayashi-Maskawa (CKM) matrix would be a strong indicator of
new physics. Two elements of this CKM matrix, $|V_\mathrm{cd}|$ and
$|V_\mathrm{cs}|$, can be deduced from the weak decays of $\rm D$ and
$\rm D_\mathrm{s}$ mesons into a lepton and a neutrino by combining
experimentally measured decay rates with theory determinations of the
decay constants $f_\mathrm{D}$ and $f_{\mathrm{D_s}}$, respectively.
Precise, ab-initio predictions of the decay constants are required and
these can be obtained using lattice QCD.  However, observables
involving both charm and light or strange quarks are challenging to
compute on the lattice with all sources of systematic uncertainty
under control. The charm quark mass does not provide a high enough
scale to admit an effective-field theory treatment, while employing a
relativistic quark action with discretisation effects typically of
$\mathrm{O}(a^2)$, including $\mathrm{O}(m_\mathrm{c}^2a^2)$, requires
simulations with fine lattice spacings. Furthermore, the light (and
strange) quark mass dependence needs to be sufficiently constrained,
in particular, close to the physical point.

In this work we present results for the D and D$_{\rm s}$ decay
constants from an analysis of $N_{\rm f}=2+1$ gauge field
configurations generated by the Coordinated Lattice Simulations~(CLS)
consortium~\cite{Bruno:2014jqa, Mohler:2017wnb, RQCD:2022xux} with
non-perturbatively $\mathrm{O}(a)$ improved Wilson
quarks~\cite{Sheikholeslami:1985ij,Bulava:2013cta} and the tree-level
Symanzik-improved gauge action~\cite{Luscher:1984xn}.  The charm quark
is introduced as a quenched flavour, thus making our setup a partially
quenched realisation of the four-flavour theory. As argued later, the
impact of a missing charm quark in the sea is expected to be below our
small total uncertainties.

A unique feature of this study is that both the continuum limit
extrapolation and light and strange quark mass dependence is tightly
constrained by the use of 49 high-statistics ensembles, which lie on
three trajectories in the (sea) quark mass plane and which span a
range of lattice spacings from $a\approx 0.10\,$fm down to below
$a\approx 0.04\,$fm ($a^2$ varies by a factor of 6). Two of the
trajectories meet at the physical point: along one trajectory the
flavour average of the light and strange quark masses is held
constant, and along the other the strange quark mass is fixed to
approximately its physical value. The third trajectory runs towards
the SU(3) chiral limit, with the light and strange quark masses set to
be equal.  Overall, the pion mass varies from $m_\pi \approx 420$\,MeV
down to 129~MeV, where for most ensembles the spatial extent $L$ is
large enough such that $m_\pi L\gtrsim 4$ and significant
finite-volume effects are avoided. Lattice spacings as fine as
$0.04\,$fm are achieved utilising open boundary conditions in
time~\cite{Luscher:2011kk} to avoid the problem of topological
freezing.

The pseudoscalar meson decay constants are defined through matrix
elements of the axial vector current between pseudoscalar states and
the vacuum. For Wilson fermions, the current is multiplicatively
renormalised and, in order to achieve leading order $\mathrm{O}(a^2)$
discretisation effects, $\mathrm{O}(a)$ improved. For the latter, this
includes evaluating terms to remove the leading quark mass-dependent
cutoff effects.  We remark that, because of the precision non-perturbative
determination of the renormalisation factor of
ref.~\cite{DallaBrida:2018tpn} and the improvement coefficients of
refs.~\cite{Bulava:2015bxa,Bali:2021qem} employed in our analysis, renormalisation and
improvement is not a significant source of uncertainty in our
results.

With both the light and strange quark masses varying
across the ensembles, we perform a simultaneous chiral and continuum
extrapolation of $f_{\rm D}$ and $f_{\rm D_s}$ with all correlations
taken into account. Any mistuning of the trajectories, where for a
particular lattice spacing the relevant trajectory does not go
through the physical point, can be corrected for using the
chiral-continuum fit parameterisation. 
Our final results at the physical point read
\begin{align}
f_{\mathrm{D_s}} &= 246.8(0.64)_{\rm stat}(0.61)_{\rm sys}(0.95)_{\rm scale}[1.3]\,\mathrm{MeV}\,, \nonumber\\
f_\mathrm{D} &= 208.4(0.67)_{\rm stat}(0.75)_{\rm sys}(1.11)_{\rm scale}[1.5]\,\mathrm{MeV}\,,\\
f_{\mathrm{D_s}}/f_\mathrm{D} &= 1.1842(21)_{\rm stat}(22)_{\rm sys}(19)_{\rm scale}[36]\,, \nonumber
\end{align}
where the first error is statistical, the second systematic~(arising
from the parameterisation of the lattice spacing and quark mass
dependence), the third component is due to the scale setting~(taken
from ref.~\cite{RQCD:2022xux}) and within the last bracket we give the total
uncertainty. These results are the most precise for $N_\mathrm{f}=2+1$
lattice QCD to-date and are consistent with the recent work of
ref.~\cite{Bussone:2023kag}, which utilises a sub-set of the ensembles
employed here (with a mixed action setup).  Our errors are more than
twice those quoted by FNAL/MILC in their $N_{\rm f}=2+1+1$
study~\cite{Bazavov:2017lyh} (see the 2021 FLAG
report~\cite{FlavourLatticeAveragingGroupFLAG:2021npn} for a
comprehensive review of lattice results). An immediate improvement in
our results would be achieved through a more precise determination of
the scale. However, once the uncertainties are reduced to the level of
a few per mille, isospin-breaking effects, as well as the absence of
charm sea quarks, will need to be considered.

The structure of the rest of the paper is as follows. In section
\ref{s:setup} we give details of our lattice setup and the ensembles
employed.  The construction of the two-point correlation functions is
discussed in section \ref{s:obs}, along with the Symanzik improvement
and renormalisation of the relevant heavy-light operators. The fitting
procedure to extract the bare decay constants from the correlation
functions is described in section \ref{s:ana}. Finite-volume effects
are not significant in our setup, and this is demonstrated in section
\ref{s:fv}.  In section \ref{s:extrap}, we outline the set of chiral
and continuum limit extrapolations performed and how the results at
the physical point are combined via a model averaging procedure. We
also specify our choice of hadronic scheme, i.e.\ the lattice scale
employed and the external (physical) input used to define the physical
quark masses. The final results are presented in section
\ref{s:results} and compared with previous works. We conclude in
section \ref{s:conc}. Additional information on the simulations,
including the masses and bare decay constants for the individual
ensembles and the scheme dependence of the final results, are collected
in appendices. Preliminary accounts of this work have been presented
in refs.~\cite{Collins:2017rhi,Collins:2017iud}.

\section{Lattice setup \label{s:setup}}

\begin{figure}[tp]
	\centering
	\includegraphics[width=0.9\textwidth]{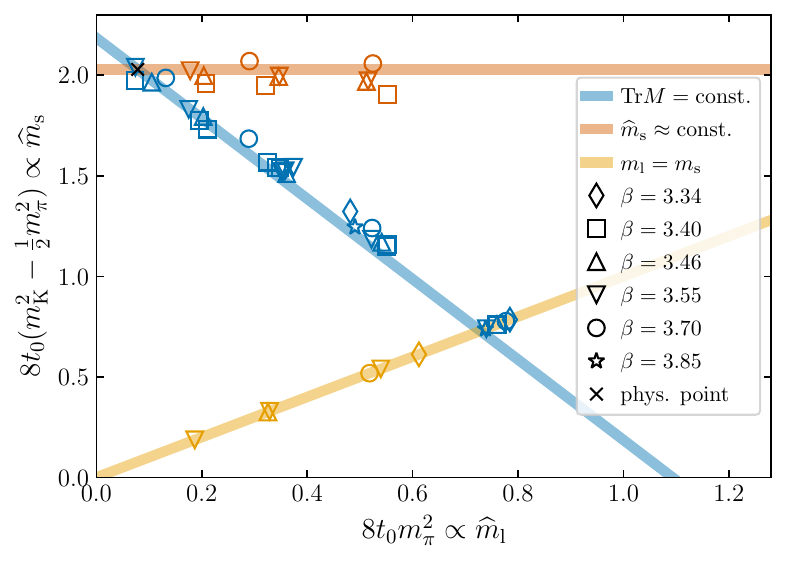}
	\caption{\label{fig:CLS}Overview of the ensembles used in this work in the plane of (renormalised) light and strange quark masses: the chiral trajectory where $\Tr M = \text{const.}$ approaches the physical point from below (blue line) and meets the trajectory with renormalised strange quark mass 
	    $\widehat{m}_{\rm s} \approx \text{const.}$ (orange line) at the physical point by
construction. A third trajectory for which the light and strange quark masses are equal, $m_{\rm l}=m_{\rm s}$ (yellow line), approaches the SU(3) chiral limit. The orange line is obtained by setting 
$8t_0 (m_{\rm K}^2 - \frac{1}{2}m_\pi^2)$ to its physical value, analogously the blue line is defined by fixing~$\Mbarsq \equiv 8 t_0 \big(2m_\mathrm{K}^2 + m_\pi^2\big) / 3= \bar{\mathbb{M}}^2_{\rm phys}$.
}
\end{figure}

In the following we describe the set of $N_{\rm f}=2+1$ ensembles utilised in our analysis. We have employed non-perturbatively $\mathrm{O}(a)$ improved Wilson fermions~\cite{Sheikholeslami:1985ij,Bulava:2013cta} and the tree-level Symanzik improved gauge action~\cite{Luscher:1984xn}. All ensembles have been created within the CLS effort.\footnote{For uptodate information on the CLS $N_{\rm f}=2+1$  ensembles, see \url{https://www-zeuthen.desy.de/alpha/public-cls-nf21/}.}
We have ensembles at six values of the squared bare coupling constant $g_0^2 = 6/\beta$, which corresponds to lattice spacings ranging from $a = 0.098\,\textmd{fm}$ down to $a = 0.039\,\textmd{fm}$. 
At each value of $\beta$, simulations are carried out along three trajectories in the quark mass plane,
as visualised in figure~\ref{fig:CLS}:
\begin{itemize}
  \item The $\Tr M=\text{const.}$ line: the trace of the (bare) quark mass matrix $M$ is kept fixed, $\Tr M  = 2m_{\rm l}+m_{\rm s}= \text{const.}$. The latter also holds for the renormalised quark masses up to $\mathrm{O}(a)$ effects, $2\widehat{m}_{\rm l}+\widehat{m}_{\rm s}=\text{const.} +\mathrm{O}(a)$. The constant
    is chosen such that the flavour average of the pseudoscalar octet meson masses squared rescaled with the Wilson flow parameter $t_0$ is close to its physical value, $\Mbarsq \equiv 8 t_0 \big(2m_\mathrm{K}^2 + m_\pi^2\big) / 3= \bar{\mathbb{M}}^2_{\rm phys}$, where $m_\pi$ and $m_\mathrm{K}$ are the masses of the pion and kaon, respectively.
    See refs.~\cite{Bruno:2016plf, RQCD:2022xux} for more details on the simulations along this trajectory.
  \item The $\widehat{m}_{\rm s} \approx \text{const}$ line: the renormalised
    strange quark mass is kept approximately constant. The constant is chosen such that $\widehat{m}_{\rm s}$ is near its physical value, $\widehat{m}_{\rm s} \approx \widehat{m}_{\rm s}^{\rm phys}$, see ref.~\cite{Bali:2016umi} for further details.
  \item The symmetric line: the light and strange quark masses are equal, $m_{\rm l}=m_{\rm s}$. This line has an intersection point with the $\Tr M=\text{const.}$ trajectory (referred to as the SU(3) symmetric point) and approaches the SU(3) chiral limit.
\end{itemize}

Note that in the literature often only either the $\Tr M=\text{const.}$ or the $\widehat{m}_{\rm s} \approx \text{const}$ trajectory is considered when chiral extrapolations are performed. In such cases, (ideally) small deviations of the simulation points from the desired chiral trajectory
arise due to discretisation effects and mistuning of the simulation parameters
(as the simulation parameters that would match
the desired trajectory can only be determined after the quark
mass plane has been sufficiently explored).
The latter deviations are often corrected by means of reweighting or a Taylor expansion~\cite{Bruno:2016plf}. The need to employ such methods is avoided in our analysis,
since our set of ensembles allows for a reliable parametrisation of
both the light and strange quark mass dependence of our observables
and the extraction of the results at the physical point.

\begin{figure}[tp]
  \centering
  \resizebox{1.\textwidth}{!}{\includegraphics[width=\textwidth]{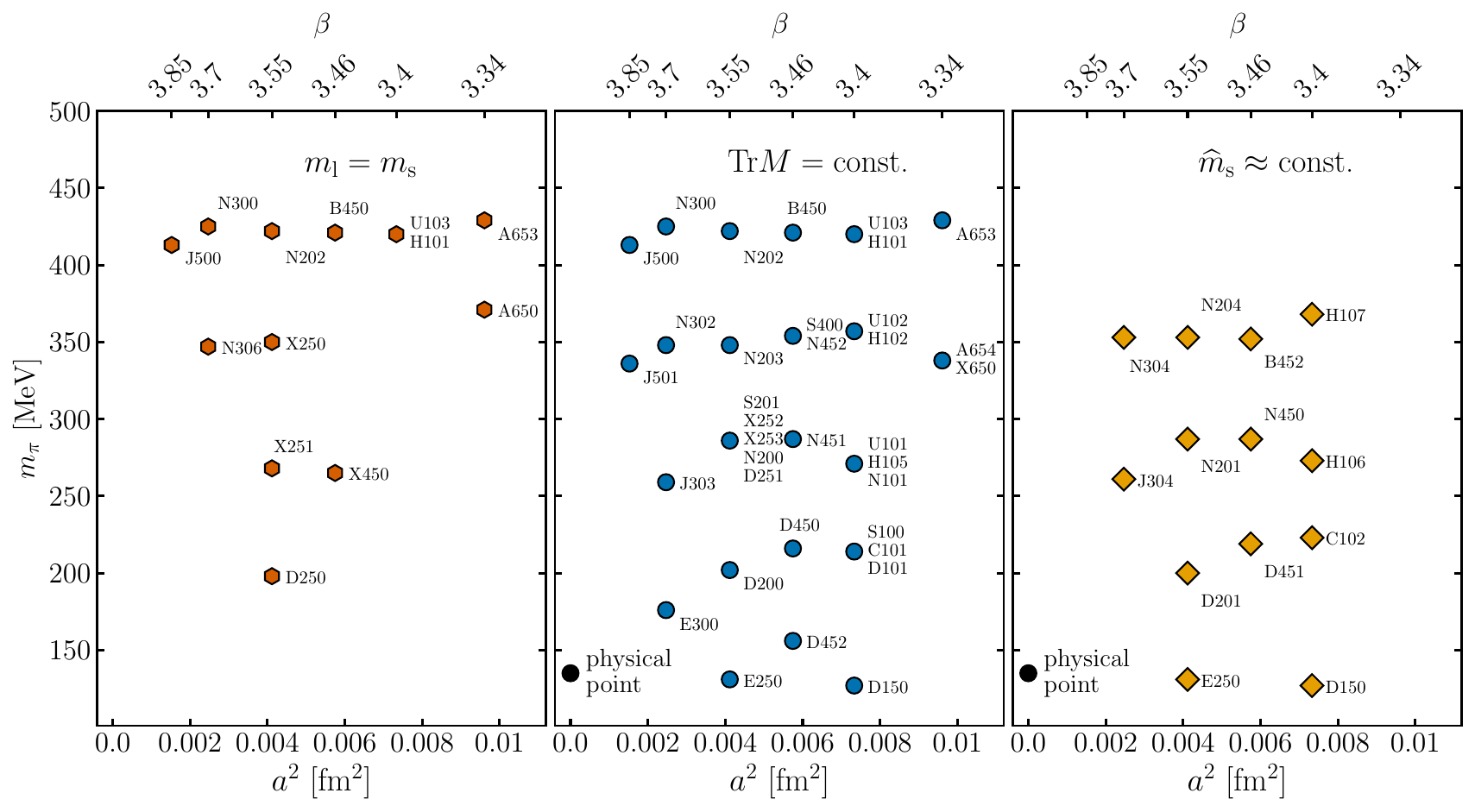}}
  \caption{
        Overview of the ensembles entering the present study: three different quark mass
    trajectories (left: $m_{\rm l}=m_{\rm s}$, centre: $\Tr M  =\text{const.}$, right: $\widehat{m}_{\rm s} \approx \text{const.}$) have been analysed at six (four for $\widehat{m}_{\rm s} \approx \text{const.}$) different lattice spacings. Additional ensembles with different spatial volumes exist for selected parameter sets. From these dedicated ensembles, the small volumes (S100, S201, U101, U102, U103, X252, X253) do not enter the final extrapolation. Also the ensembles B452 and N300 have been excluded. For these two cases the rather small temporal extent, in view of the given statistics, did not allow for a reliable extraction of the ground state.
\label{fig:ensembles}}
\end{figure}

An overview of the lattice spacings and pion masses of the ensembles
we utilise is shown in figure~\ref{fig:ensembles}. More details are
given in~\cref{tab:overview,t:hopping} in appendix~\ref{a:tables}.  In
order to circumvent critical slowing down towards the continuum limit
due to the freezing of the topological charge~\cite{Schaefer:2010hu},
open boundary conditions in time~\cite{Luscher:2011kk} have been
implemented for ensembles with $a\lesssim 0.05\,$fm, while for 
other ensembles, antiperiodic boundary conditions have also been used. In
the spatial directions, periodic boundary conditions are always
imposed. We emphasise that for four of the six lattice spacings (the
exceptions are the coarsest and finest lattice spacings) the ensembles provide
excellent coverage of the quark mass plane in the region of
interest. All together, this setup enables a detailed investigation of
the quark mass and cutoff dependence when performing combined
chiral-continuum extrapolations  of the decay constants to the physical point.

The charm quark is partially quenched in our analysis.  Two values of
the charm quark mass are employed per ensemble, which are chosen such
that only a small interpolation or (in a few cases) extrapolation to
the physical value is required.
The dependence of the charm quark mass on the light and strange sea quark masses is expected to be mild
and we usually use the same hopping parameters for a given gauge coupling. The values of the hopping parameters for the light, strange and charm quarks are listed in~\cref{t:hopping}.

Finite-size effects can also be quantified 
in our analysis. Ensembles were generated with different spatial
extents at several simulation points for $m_\pi>200$\,MeV
and $a\ge 0.064\,$fm. In particular, at $m_\pi \approx 286$\,MeV and $a=0.064\,$fm, five
volumes are realised with $m_\pi L$ ranging from 3.0 to 5.9.  As
detailed in \cref{s:fv}, we find that finite volume effects are under
control for $m_\pi L\gtrsim 3.5$. We impose an additional cut of
$L\ge 2.3\,$fm to ensure that the spatial extent is much larger than the
inverse pseudo-critical temperature. We remark that for almost all simulation points there
exists at least one ensemble with a spatial lattice extent $L
\gtrsim 4/m_\pi$.

We achieve high statistics across our set of ensembles.  In
particular, for almost all ensembles we have a large number of
configurations with respect to the (possibly) slowest mode in the
simulation.  See \cref{tab:overview,} of appendix~\ref{a:tables} for
the number of configurations utilised for each ensemble and
ref.~\cite{RQCD:2022xux} for details on the autocorrelations. The
resulting precision of better than $0.6$\textperthousand\,  and
6\textperthousand\, for the heavy-light meson masses and bare decay
constants, respectively, is illustrated in~\cref{t:ps,t:decayc}.

More details on the simulations performed by CLS
using the {\sc openQCD} code~\cite{Luscher:2012av, openQCD}
can be found in refs.~\cite{Bruno:2014jqa, Mohler:2017wnb, RQCD:2022xux}; we highlight here that the code package has several algorithmic improvements built in, such as the Hasenbusch trick~\cite{Hasenbusch:2001ne}, improved integrators~\cite{Omelyan2003272}, a multi-level integration scheme~\cite{Sexton:1992nu} and a deflated solver~\cite{Luscher:2007es,Frommer:2013fsa}. Furthermore, for the $N_{\rm f}=2$ light fermion part of the action, a twisted mass term is added in order to increase the stability of the HMC simulation~\cite{Luscher:2012av, Kuberski:2023zky}. The effect of this additional term is removed by appropriate reweighting of the observables. Note that reweighting is also applied in the strange sector in order to account for errors arising from the use of a rational approximation~\cite{Kennedy:1998cu,Clark:2006fx}. The strange quark reweighting factor usually does not vary significantly, however, it turns out that the factor can acquire a negative sign~\cite{Mohler:2020txx}, which should be included in the ensemble averages. We utilise the signs calculated in refs.~\cite{Mohler:2020txx, Kuberski:2023zky} in our analysis. Only very few configurations have a negative strange reweighting factor and the effect of the negative signs is to (at most) slightly increase the statistical uncertainty.

\section{Observables \label{s:obs}}

The pseudoscalar decay constants $f_{\mathrm{D}}$ and $f_{\mathrm{D}_{\rm s}}$ are defined via the matrix elements of the axial vector current between $\mathrm{D}$ and $\mathrm{D}_{\rm s}$ meson states at momentum $p$ and the vacuum, respectively,
\begin{equation}
 \big< 0\big\vert A^{\rm lc}_\mu\big\vert \mathrm{D}(p)\big> =
 \mathrm{i} f_{\mathrm{D}} p_{\mu}, \qquad
 \big< 0\big\vert A^{\rm sc}_\mu\big\vert \mathrm{D}_{\rm s}(p)\big> =
 \mathrm{i} f_{\mathrm{D}_{\rm s}} p_{\mu}.
 \label{eq:Fps}
\end{equation}
The axial vector current is given by $A^{q \rm c}_{\mu}(x) = \overline{q}(x)\gamma_\mu\gamma_5 c(x)$ for quark flavours ${q}={\rm l},{\rm s}$. In order to remove $\mathrm{O}(a)$ cutoff effects from the matrix elements,
we construct an improved current 
\begin{equation}
 A^{q \rm c,I}_{\mu}=
 A^{q \rm c}_{\mu}
 +a c_{\rm A} \frac{1}{2}\left(\partial_\mu+\partial^*_\mu\right)P^{q\rm c}, \label{eq:impc}
\end{equation}
where the pseudoscalar operator is $P^{q \rm c}(x) = \overline{q}(x)\gamma_5 c(x)$
and $\partial_\mu$ and $\partial^*_\mu$ denote the lattice forward and
backward derivatives, respectively.  Together with an $\mathrm{O}(a)$ improved
fermion action, the above current eliminates all $\mathrm{O}(a)$
effects in the chiral limit.  At non-vanishing quark mass, two
additional terms are needed which depend on the valence quark masses
and the sum of the sea quark masses.  The renormalised $\mathrm{O}(a)$ improved
 current reads \cite{Bhattacharya:2005rb}
\begin{equation}
 \left(A^{q\rm c,I}_{\mu}\right)_{\rm R}=
 Z_{\rm A}\left[1+a\left(b_{\rm A} m_{q\rm c}+ \bar{b}_{\rm A} \Tr M\right)\right]
 A^{q\rm c,I}_{\mu}  +{\mathrm O}(a^2),
 \label{eq:RenImp}
\end{equation}
where $m_{q\rm c}$ and $\Tr M$ denote the bare vector Ward identity quark mass
combinations
\begin{equation}
 m_{q \rm c}=
 \frac{1}{2}\left(m_{q}+m_{\rm c}\right), \quad \Tr M  = 2m_{\rm l}+m_{\rm s},  \quad \text{with } m_{q} = \frac{1}{2 a} \left(\frac{1}{\kappa_{q}} - \frac{1}{\kappa_{\rm crit}}\right).
 \label{vectorWardIdentity}
\end{equation}
The hopping parameter for quark flavour $q$ is denoted by $\kappa_{q}$
and $\kappa_{\rm crit}$ labels its critical value. For the
renormalisation factor  $Z_{\rm A}$ and the improvement coefficients $c_{\rm A}$ and $b_{\rm
  A}$, we employ the
non-perturbative determinations of refs.~\cite{DallaBrida:2018tpn,
  Bulava:2015bxa, Bali:2021qem}, respectively.  The improvement coefficient $\bar{b}_{\rm A}$
has been computed in refs.~\cite{Bali:2021qem,Korcyl:2016ugy},
however, as the coefficient is compatible with zero for the range of
gauge couplings considered here, we set $\bar{b}_{\rm A}=0$ in our
analysis. For $\kappa_\mathrm{crit}$ we utilise the results of
ref.~\cite{RQCD:2022xux}.  For convenience, we collect the values for the renormalisation factor, improvement
coefficients and $\kappa_\mathrm{crit}$ for each
gauge coupling used in this work in~\cref{tab:inputparam} of
appendix~\ref{a:tables}.

We remark that the gauge coupling $g_0^2$ also undergoes $\mathrm{O}(a)$
improvement, $\tilde{g}_0^2 = g_0^2 (1 + \frac{1}{3} a b_g(g_0^2) \Tr
M)$, where to consistently apply Symanzik improvement to this order,
the improved coupling should be held constant away from the chiral
limit~\cite{Luscher:1996sc,Bhattacharya:2005rb}. While there has been
a recent effort to calculate $b_g$
non-perturbatively~\cite{DallaBrida:2023fpl}, there is no
determination available as yet for the range of coupling constants
relevant for our calculations and the simulations are performed at
fixed values of $g_0$. Note that the renormalisation factor
and coefficients of the mass-dependent improvement terms
in eq.~\eqref{eq:RenImp} should be evaluated at $\tilde{g}_0^2$. 
However, for the latter the difference between, for instance, $b_{\rm A}$ evaluated at
$g_0^2$ or $\tilde{g}_0^2$ enters at $\mathrm{O}(a^2)$ and can be ignored.
To leading order in perturbation theory $b_g(g_0^2) =
0.036g_0^2 + \mathrm{O}(g_0^4)$~\cite{Sint:1995ch}. Using this
estimate, the values of the improved coupling associated with our
ensembles are very close to $g_0^2$ and in practice we evaluate renormalisation
factors (as well as the improvement coefficients) at the gauge coupling where we have actually performed the
simulation.

Holding $g_0^2$ (and not $\tilde{g}_0^2$) fixed in the simulations,
means that the lattice spacing depends (mildly) on the quark masses,
$a=a(g_0^2,a \Tr M)$. This introduces (mass-dependent) cutoff effects
of $\mathrm{O}(a)$ in dimensionful quantities extracted in the
simulation~(at fixed $g_0^2$), such as a meson mass $am_{\rm M}$. This effect
is also expected to be small.  Nevertheless, we circumvent this issue
by rescaling all dimensionful quantities with appropriate powers of
the gradient flow parameter $t_0$ (determined on the same ensemble) to
form dimensionless combinations such that the $\mathrm{O}(a)$ effects cancel.

In order to obtain the matrix elements of~\cref{eq:Fps}, we evaluate the two-point functions
\begin{alignat}{2}
 C_{A_0\tilde{P}}^{q\mathrm{c}}(t) &\equiv C_{A_0\tilde{P}}^{q\mathrm{c}}(x_0, y_0)&&=
 -\frac{a^6}{L^3}\sum_{\vec{x},\vec{y}}
 \Big< A_0^{q\rm c,I}(x)\left(\tilde{P}^{q\rm c}(y)\right)^\dagger\Big>,  \nonumber \\
  C_{\tilde{P}\tilde{P}}^{q\mathrm{c}}(t) &\equiv C_{\tilde{P}\tilde{P}}^{q\mathrm{c}}(x_0, y_0)&&=
 -\frac{a^6}{L^3}\sum_{\vec{x},\vec{y}}
 \Big< \tilde{P}^{q\rm c}(x)\big(\tilde{P}^{q\rm c}(y)\big)^\dagger\Big>,
 \label{eq:corrs}
\end{alignat}
at zero momentum, where $t=x_0-y_0$ is the difference between the sink and source timeslices, $x_0$ and $y_0$, respectively. The spatial extent is denoted by $L$. The correlators are calculated by means of point-to-all propagators, where, for the pseudoscalar interpolator at the source and sink $\tilde{P}^{q\rm c (\dagger)}$, we apply Wuppertal smearing~\cite{Gusken:1989ad,Gusken:1989qx} with APE-smoothed links~\cite{Falcioni:1984ei}.
The number of smearing iterations is varied across the ensembles to optimise the overlap with the ground state. More iterations are needed as the light quark mass decreases and we rescale the number of iterations with $a^{-2}$ to ensure (roughly) the same root mean square quark smearing radius as the lattice spacing is varied. In~\cref{tab:overview} of appendix~\ref{a:tables}, the individual numbers of smearing iterations for each quark  are summarised for all ensembles. For more details related to the smearing, we refer the reader to ref.~\cite{RQCD:2022xux}.

We compute the above two-point functions for two charm quark masses close to the physical charm quark mass. For a given gauge coupling, the same two values of the hopping parameter are normally utilised for each ensemble and the number of smearing iterations employed for the charm quark is also kept fixed.
Note that the charm propagators are computed with the highest numerical precision that is possible with our code: we impose relative residuals of around $10^{-15}$. No problems due to the numerical accuracy of the charm quark propagators are observed when fitting to the correlation functions to extract the heavy-light meson masses and decay constants (within the fit ranges chosen). Our fitting procedure is outlined in the following section.

In order to increase statistics, source operators are inserted at different
temporal positions. For ensembles with open boundaries, the source
positions are placed on a fixed set of time slices (equally distributed
within the bulk region, with random spatial positions), while for
those with periodic boundaries, the source positions are chosen
randomly for each configuration.
The total number of sources (per configuration) for each ensemble can be found in table~\ref{tab:overview}.

\section{Extraction of the bare quantities \label{s:ana}}
In this section we describe how we extract the masses and decay
constants of the heavy-light and heavy-strange mesons from fits to the
relevant two-point correlation functions. In the first step, we
average the two-point functions over the source positions. While this
is straightforward on ensembles with periodic boundary conditions,
care has to be taken when open boundaries in the time direction are
imposed as the data can only be averaged in the bulk of the lattice,
where boundary effects are negligible and translational invariance is effectively restored.
We identify the boundary region by fitting the $\tilde{P}\tilde{P}$ 
correlation function to the form
suggested in ref.~\cite{Luscher:2012av}, which accounts for states which propagate
in from the boundary, and by applying the criterion that the contribution of these states is less than
$\textstyle\frac{1}{4}$ of the statistical uncertainty of the
correlation function.
To be conservative, a minimal distance of $1.5\,$fm is taken, even when boundary states
have decayed earlier.  For the heavy-light mesons, this is the case
for almost all ensembles.  We then average over all two-point functions where
the source and sink are in the bulk region. 
The resulting correlation function is free from significant boundary effects and only depends on the
source-sink separation, although contributions from excited states are still present.

We take a similar approach when extracting the ground state masses and matrix elements of interest. We determine
the region where excited state contamination has fallen below the noise
by fitting the effective mass of each correlation function with the two-state functional form,\footnote{We define the effective mass of a correlation function $C(t)$ by 
	$m_{\rm eff}(t) = \log[C(t - a) / C(t + a)]/(2a)$ in the case of open
	boundary conditions, and in the case of periodic
	boundary conditions as the solution of $C(t) / C(t+1) = \cosh[m_{\rm eff}(t) \cdot
	(t - T/2)] / \cosh[m_{\rm eff}(t) \cdot (t + 1 - T/2)]$.}
\begin{align}
	m_{\rm eff}(t) = m_0 + c_1 \exp(-m_1 t)\,,
\end{align}
over the range $[t_{\rm min},t_{\rm max}]$ and using the criterion from 
above.
The upper end of the fit region $t_{\rm max}$ is set by
the time slice where the relative statistical uncertainty of the effective
mass exceeds $8\%$.
To fix the start of the fit region $t_{\rm min}$, we scan over various
fits, increasing $t_{\rm min}$ while keeping $t_{\rm max}$ fixed. The best fit
is identified by monitoring the fit quality, as given by the reduced $\chi^2$
for uncorrelated fits introduced in ref.~\cite{Bruno:2022mfy}. To avoid
overfitting in a region where only a single state can be resolved, we then 
identify the best fit to be the one with the highest Akaike weight, computed as outlined in section
\ref{s:model}. In general, we find the determination of the region of ground state
dominance to be largely insensitive to the procedure used to choose the
best fit.
The resulting regions of ground state dominance start in the range from 
0.6\,fm to 1.6\,fm, where the variation within this range depends on the 
quark masses and the smearing parameters, as well as on the statistical 
accuracy of the data.

\begin{figure}[t]
	\centering
	\includegraphics[width=0.5\textwidth]{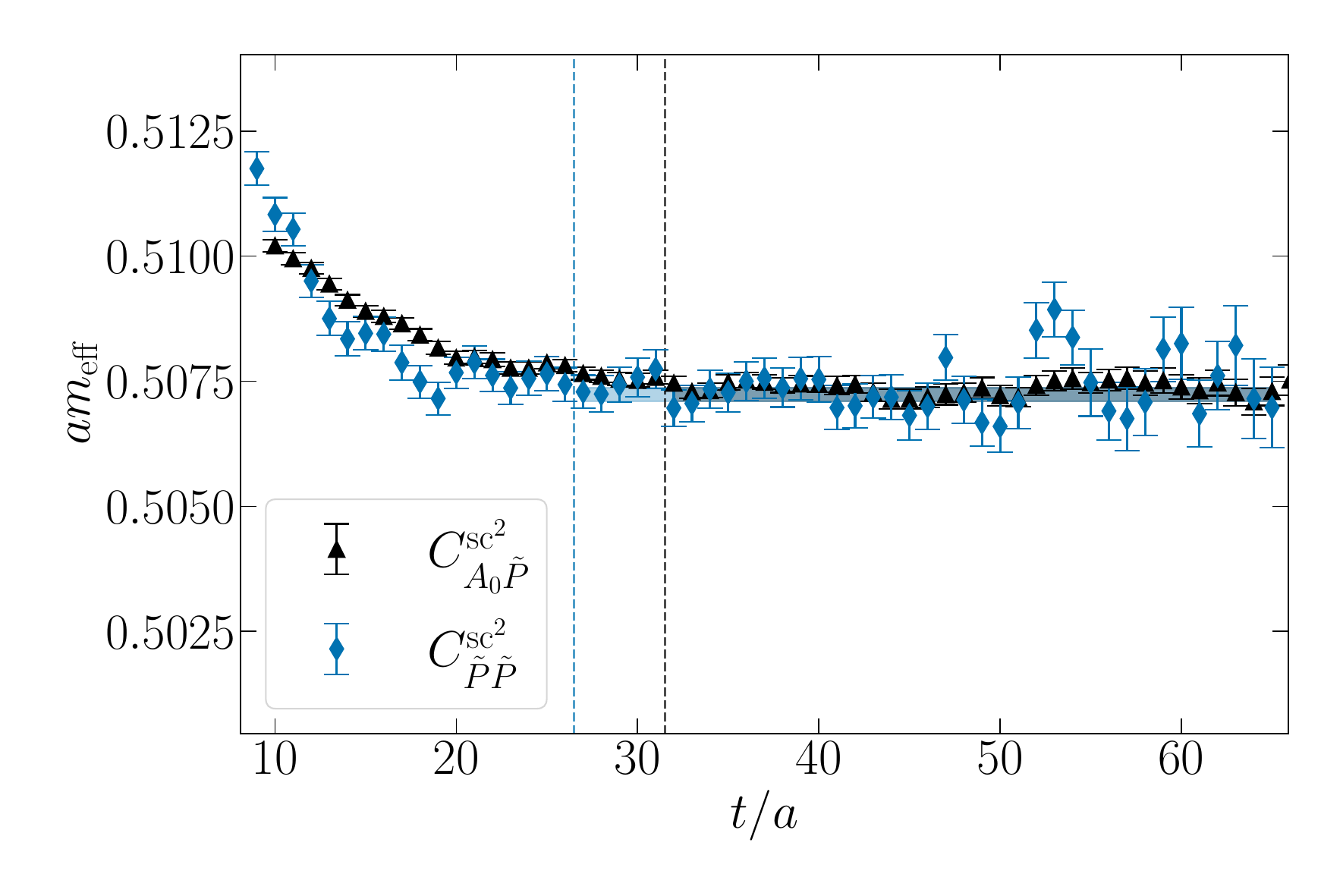}%
	\includegraphics[width=0.5\textwidth]{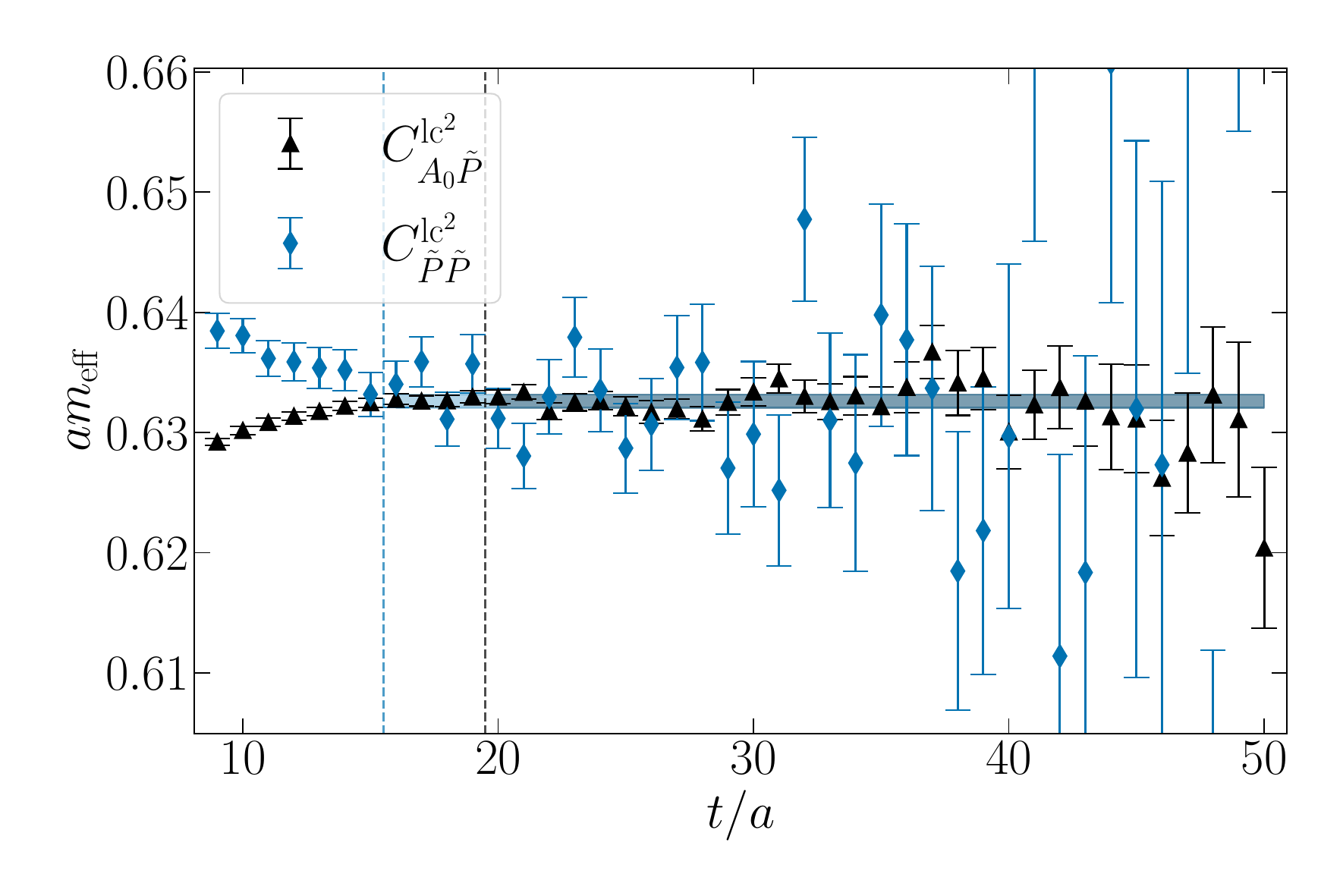}\\
	\includegraphics[width=0.5\textwidth]{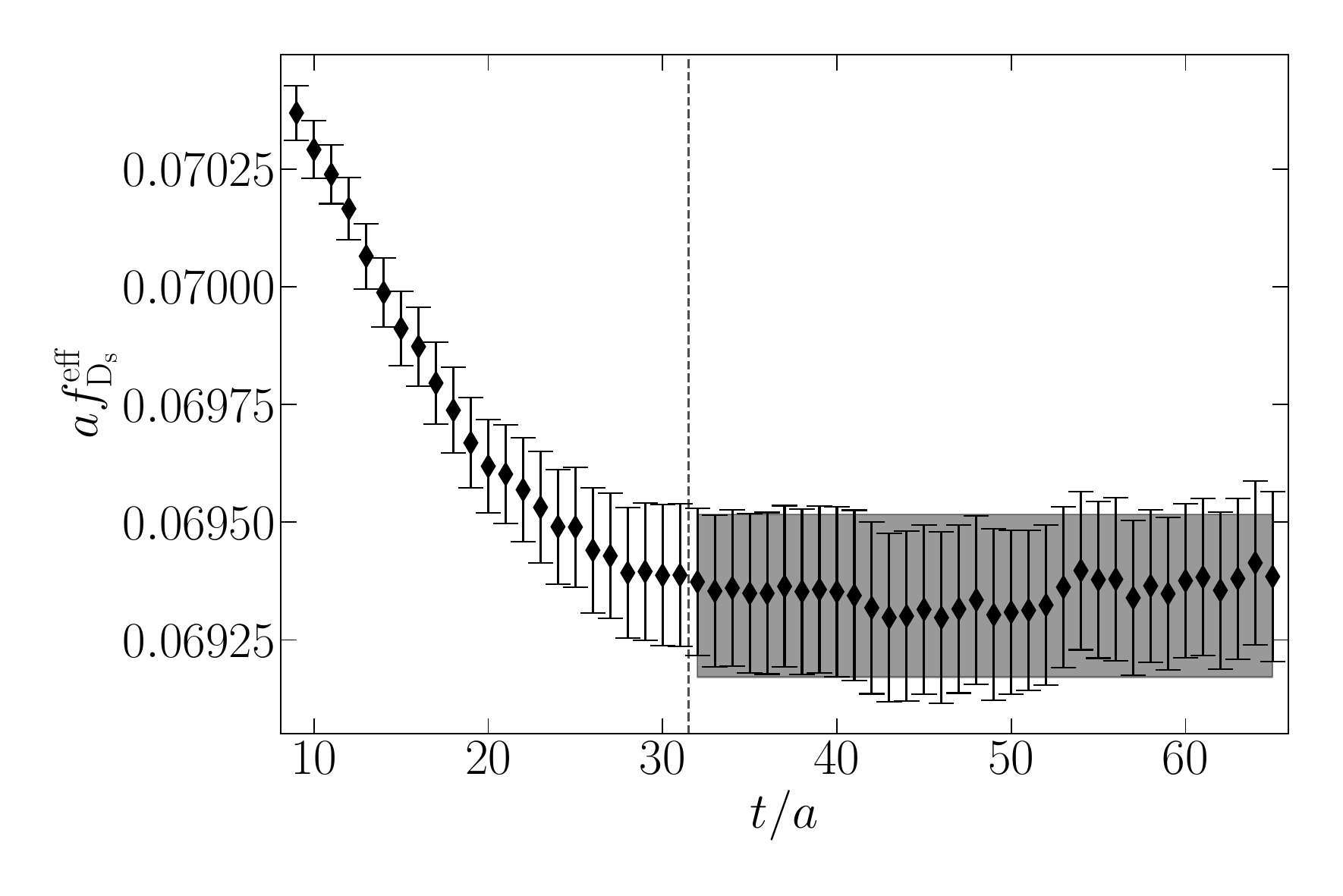}%
	\includegraphics[width=0.5\textwidth]{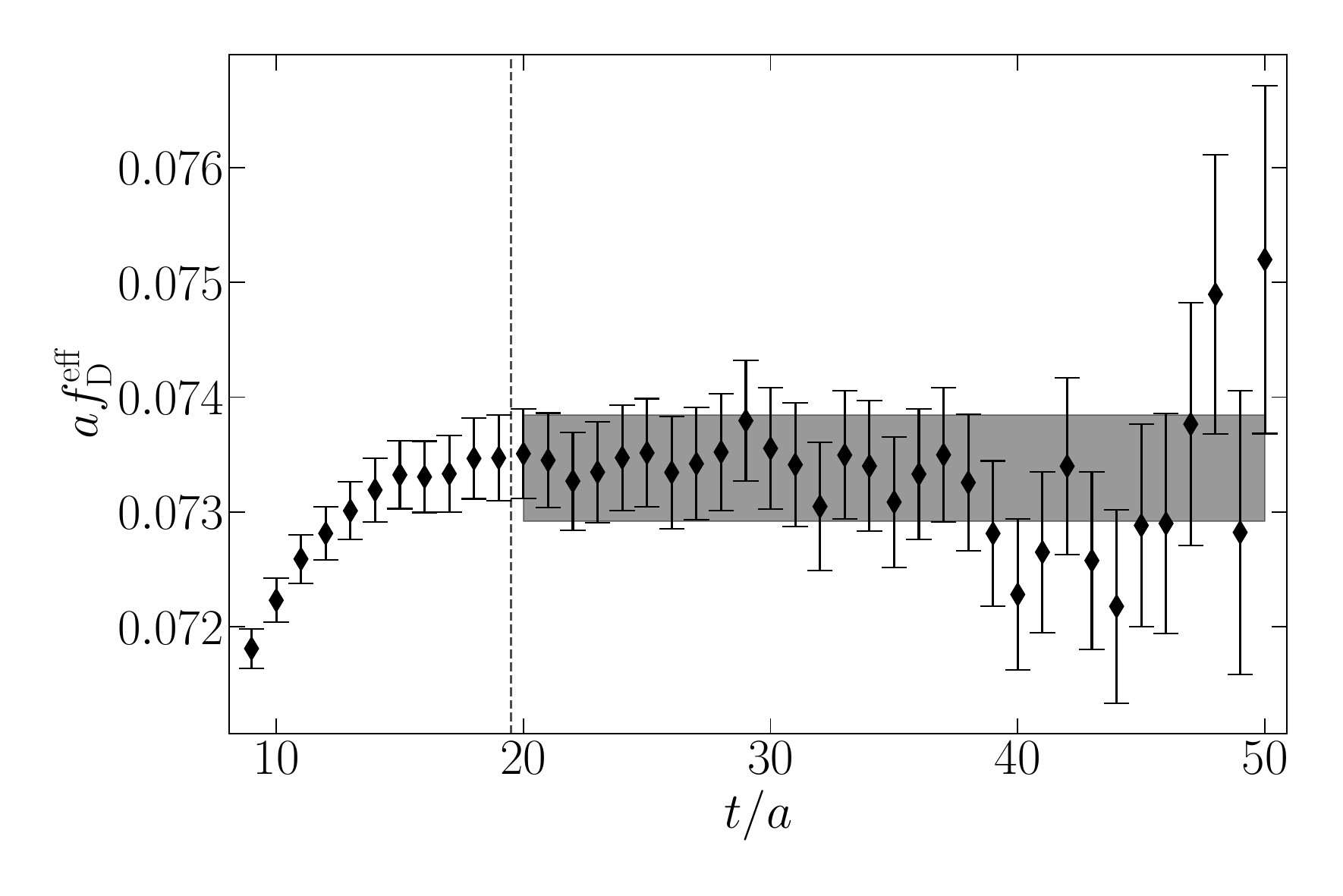}%
	\caption{Results from our combined fits to the $\tilde{P}\tilde{P}$ and $A_0\tilde{P}$ two-point functions for the heavy-strange meson on ensemble E300 (left, $m_\pi=175$\,MeV, $a=0.049\,$fm) and the heavy-light meson on ensemble E250 (right, $m_\pi=130$\,MeV, $a=0.064\,$fm).   Note that the results displayed are for the second charm hopping parameter ($\kappa_{c^2}$), see table \ref{t:hopping} in appendix \ref{a:tables}.
	{\it Top}: The fit result for the meson mass $m_{\mathrm{D}_{(\rm s)}}$ (blue band) is shown together with the effective masses of the two correlation functions. The lower end of the fit range~($t_{\rm min}$) in each case is indicated by a dashed line. {\it Bottom}: The result for the decay constant obtained from the fit (grey band) is displayed along with the effective decay constant (eq.~\eqref{eq:efff}). The dashed line indicates the $t_{\rm min}$ of the $A_0\tilde{P}$ two-point function.
	}
	\label{f:simfit}
\end{figure}

In the next step, we perform a combined fit to the
$\tilde{P}\tilde{P}$ and $A_0\tilde{P}$
correlation functions within the fit ranges determined above.
Excited state contributions can be neglected and the correlation functions as defined in~\cref{eq:corrs}
are modelled with single-exponentials with the same energy
$m_{\mathrm{D}}$ for $q=\rm l$, and $m_{\mathrm{D}_{\rm s}}$ for $q=\rm s$
and amplitudes $A_{A_0\tilde{P}}^{q\mathrm{c}}$ and $A_{\tilde{P}\tilde{P}}^{q\mathrm{c}}$,
\begin{align}
	C_{A_0\tilde{P}}^{q\mathrm{c}}(t)
	=
	A_{A_0\tilde{P}}^{q\mathrm{c}}\,\mathrm{e}^{-m_{\mathrm{D}_{(\rm s)}}t}\,,\qquad 
	C_{\tilde{P}\tilde{P}}^{q\mathrm{c}}(t)
	=
	A_{\tilde{P}\tilde{P}}^{q\mathrm{c}}\,\mathrm{e}^{-m_{\mathrm{D}_{(\rm s)}}t}\,.
\end{align}
Note that the $A_0\tilde{P}$
correlation function is constructed using the improved axial vector current in eq.~\eqref{eq:impc}.
The combined fit helps to reduce the
statistical uncertainty of the smeared-smeared $\tilde{P}\tilde{P}$ amplitude
that is inherently noisier. 
Based on the spectral decomposition of the correlators,
the bare matrix elements are extracted from the ground state energy and amplitudes via
\begin{align}
	\label{e:decayc_bare}
	f_{\mathrm{D}_{(\rm s)}}=\frac{\sqrt{2}A_{A_0\tilde{P}}^{q\mathrm{c}}}{\sqrt{A_{\tilde{P}\tilde{P}}^{q\mathrm{c}} m_{\mathrm{D}_{(\rm s)}}}}\,. 
\end{align}
The renormalisation and remaining (mass-dependent) $\mathrm{O}(a)$ improvement are applied as outlined in section \ref{s:loss}.

Representative results of the combined fits are displayed in
figure~\ref{f:simfit}.  In the upper panels the fitted masses of the
heavy-strange meson on ensemble E300 ($a = 0.05\,$fm, $m_{\rm K}
= 496\,$MeV) and the heavy-light meson on ensemble E250
($a = 0.064\,$fm, $m_\pi \approx m_\pi^{\rm phys}$), are shown
together with the effective masses of the corresponding
$\tilde{P}\tilde{P}$ and $A_0\tilde{P}$
correlators. One can clearly see that the extraction of the mass
is mostly constrained by the smeared-point two-point
function. Thanks to the smearing of the pseudoscalar
interpolator at the source, we can easily identify a region where excited
state contributions are sufficiently suppressed. The lower panels compare the results for the decay
constant, computed as given in \cref{e:decayc_bare}, with an effective
decay constant constructed from the correlation functions and
the fitted meson mass
\begin{align}
	f_{\mathrm{D}_{(\rm s)}}^{\rm eff}(t)
	=
	\frac{\sqrt{2} C_{A_0\tilde{P}}^{q\mathrm{c}}(t)}{\sqrt{C_{\tilde{P}\tilde{P}}^{q\mathrm{c}}(t) m_{\mathrm{D}_{(\rm s)}} \exp(- m_{\mathrm{D}_{(\rm s)}} t) }}\,.
        \label{eq:efff}
\end{align}
These figures demonstrate that there are no significant excited state contributions 
present in the region, where we extract the meson masses and decay constants.
For completeness, the analogous results for the heavy-light meson for ensemble E300 and the heavy-strange
meson for ensemble E250 are displayed in figure \ref{f:simfit_extra} in appendix \ref{a:plots}.

We collect our results for the bare decay constants from $\mathrm{O}(a)$
improved currents for both choices of the 
heavy quark mass, enclosing the physical charm for almost all ensembles, in table~\ref{t:decayc} of appendix~\ref{a:tables}. We find a statistical precision
at the level of about $0.5\%$ in most cases. 
The statistical uncertainties and the (auto-)correlations of the Monte Carlo
data are determined and propagated using the $\Gamma$-method~\cite{Wolff:2003sm,Ramos:2018vgu}. We utilise
the \texttt{pyerrors} package implementation of this method~\cite{Joswig:2022qfe}. An analysis using the
bootstrap method has also been performed as a consistency check.
For the simultaneous interpolation to the physical point and
extrapolation to the continuum limit, detailed in
section~\ref{s:extrap}, the masses of the light-light, light-strange
and heavy-heavy mesons are also needed along with the flow scale
$t_0$.  The meson masses are extracted adopting the same procedure as outlined above. 
However, for the determination of the boundary for the pion, the smeared-source local-sink correlation function
has been used (instead of the $\tilde{P}\tilde{P}$ correlation function), 
which for this particular case leads to a more conservative result. 
Table~\ref{t:ps} in appendix \ref{a:tables} lists the meson masses
and $t_0$ in lattice units obtained on each ensemble.
Note that, in addition to the ensembles with small volumes discussed in the next section, ensembles 
B452 and N300 have been also excluded. For these two ensembles, the rather small temporal extent in view of the given statistics did not allow for a reliable extraction of the ground state.

\section{Finite-volume effects \label{s:fv}}
\begin{figure}[t]
\centerline{
	\includegraphics[width=0.5\textwidth]{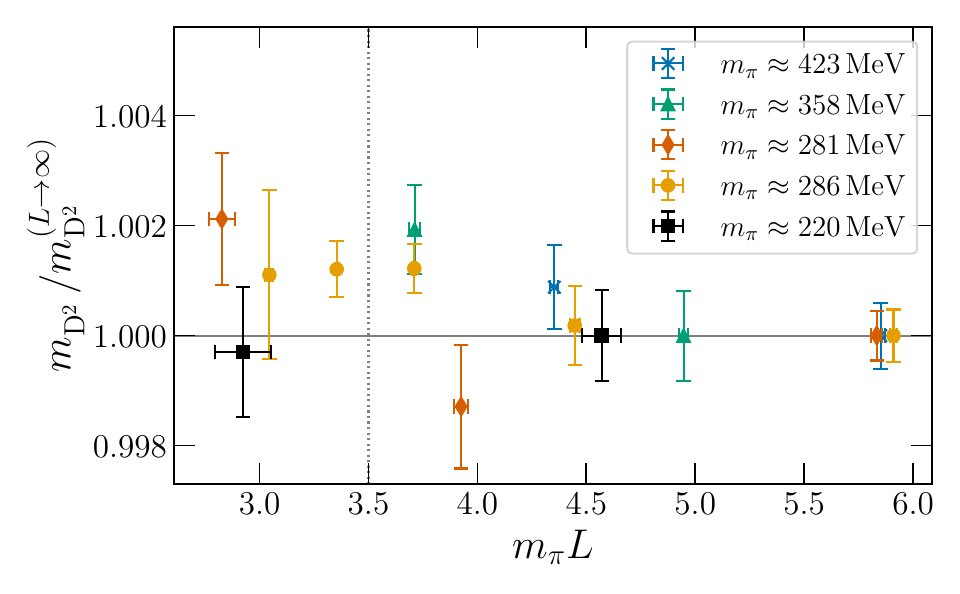}
	\includegraphics[width=0.5\textwidth]{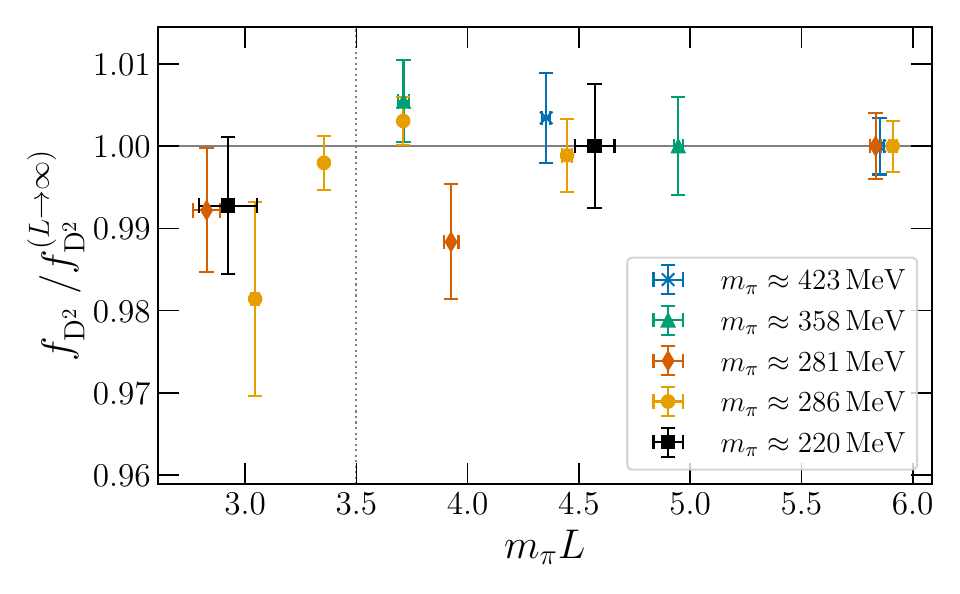}}
	\caption{Volume dependence of the heavy-light meson mass~(left) and decay constant~(right). Results for fixed gauge coupling and light and strange quark masses but different spatial extents are normalised to the value on the ensemble with the largest spatial volume. Ensembles with $m_\pi L < 3.5$ or $L<2.3\,$fm are excluded from the final analysis resulting in the omission of the six leftmost data points and the blue data point around $m_\pi L\approx 4.5$. Note that the ratios are formed using the results for the second charm hopping parameter ($\kappa_{c^2}$), see table \ref{t:hopping} in appendix \ref{a:tables}. For the $m_\pi\approx 286$\,MeV ensembles, the lattice spacing is $a=0.064\,$fm, while for all others $a=0.085\,$fm.
	\label{f:fvc}}
\end{figure}

In this analysis, our goal is to ensure that all systematics are under
control, including effects arising from simulating on a finite spatial
volume.  The latter are expected to be small for heavy meson
observables, such as the mass~\cite{Colangelo:2010ba} and decay
constant. We test this expectation by analysing ensembles with the
same light and strange quark masses and gauge coupling but different
spatial extents. The volume dependence of $m_{\rm D}$ and $f_{\rm D}$
for five sets of ensembles is shown in figure~\ref{f:fvc}. These
ensembles cover a range in the pion mass from $m_\pi\approx 423$\,MeV
down to 220~MeV and in the spatial extent from 2.1\,fm up to 4.1\,fm, with
$2.8\le m_\pi L \le 5.9$. In particular, for $m_\pi\approx 286$\,MeV,
we have results for five different values of $L$. 
Modulo fluctuations in the data, the heavy-light meson mass
tends to increase as $m_\pi L$ falls below 4, while no clear dependence can be resolved for the decay constant, although we cannot exclude a small decrease for $m_\pi L \lesssim 3$. To be conservative, we omit all ensembles with $m_\pi L < 3.5$ or $L < 2.3\,$fm, namely ensembles S100, S201, U101, U102, U103, X252, X253. Among the remaining 40 ensembles, no dependence on the volume within uncertainties is found. This also holds for the heavy-strange meson masses and decay constants and light-light and light-strange meson masses, which also enter the analysis. Therefore, we decide not to perform an explicit infinite-volume extrapolation of the remaining data. For completeness, the volume dependence of
$m_{\rm D_s}$, $f_{\rm D_s}$, $m_\pi$ and $m_{\rm K}$ is given in figure \ref{f:fvc_extra} in appendix \ref{a:plots}.

\section{Chiral and continuum extrapolation \label{s:extrap}}
To determine the decay constants at the physical point of isospin
symmetric, $2+1$ flavour QCD, we extrapolate to the continuum limit
and evaluate the decay constants for physical light, strange and charm
quark mass.  In our setup the approach to the continuum is controlled
by ensembles with larger-than-physical light quark masses at six
values of the lattice spacing, whereas 
the light and strange quark mass dependence is tightly constrained at the physical point
and the SU(3) symmetric point 
by ensembles which lie on three trajectories in the quark mass plane~(see
section~\ref{s:setup} for more details).
We perform a combined fit to the heavy-light and heavy-strange decay
constants, which by construction are equal on the ensembles with SU(3)
symmetry. Flavour symmetry provides further strong restrictions on the
description of the data.

The charm quark is partially quenched in our setup.  On each
ensemble, the charmed observables are evaluated for two values of the
heavy quark mass that encompass the physical charm quark mass. We
choose not to interpolate our data to the physical point
ensemble-by-ensemble but instead to fit all the data together and to
account for the heavy quark mass dependence in the fit
parametrisation. As there are 80 data points in total in the vicinity of the charm quark mass,
this enables us to go beyond a simple linear interpolation.

\subsection{Fixing the hadronic scheme \label{s:scheme}}
External input is required to set the lattice spacing in physical units,
as well as to adjust the simulated masses of the light, strange and charm
quarks to match those in nature.
As described in succeeding subsections, this is realised by evaluating fits to the
lattice spacing and meson mass dependence of the decay constants at
the physical point, which therefore must be specified and defines our hadronic
scheme.
For scale setting, we employ the physical value of the gradient flow
scale~\cite{Luscher:2010iy}, $\sqrt{t_0^{\rm phys}} = 0.1449^{(7)}_{(9)}\,$fm,
determined utilising the $\Xi$ baryon mass from ref.~\cite{RQCD:2022xux}.
In the light quark sector, we define the physical point in isospin symmetric
QCD using the values for the pion and kaon masses given in FLAG's 2016
review~\cite{Aoki:2016frl},
\begin{align}\label{eq:physval}
m_\pi = 134.8(3)\,{\rm MeV}\,,\qquad m_{\rm K} = 494.2(3)\,{\rm MeV}\,.
\end{align}

In the heavy quark sector, one would ideally fix the (valence) charm quark
mass using a quantity that can be determined very accurately and that is
largely insensitive to the dynamical light and strange (sea) quark content
of the ensembles.
One such quantity is the mass of the pseudoscalar charmonium meson.
This choice introduces a small intrinsic imprecision, because in our
analysis we neglect the quark-line disconnected contributions to the
corresponding mass-degenerate pseudoscalar correlation function used to
extract the charmonium mass, and likewise QED effects are omitted.
To account for this, for the physical $\eta_{\rm c}$ meson mass we adopt the
result $m_{\eta_{\rm c}}=2978.3(1.1)\,$MeV quoted in
ref.~\cite{Colquhoun:2023zbc} and originally calculated in
ref.~\cite{Hatton:2020qhk}.
It applies to pure $N_{\rm f}=2+1+1$ QCD, i.e.\ in the absence of
quark-disconnected diagrams and QED effects, and thus complies with our
situation especially in these respects. 
Compared to the experimental value,
$m_{\eta_{\rm c}}^{\rm exp} = 2983.9(4)\,$MeV~\cite{ParticleDataGroup:2022pth}, there is a $5.6\,$MeV shift such that we advocate as $\eta_{\rm c}$
input, fixing the charm quark mass, the central value
from~\cite{Colquhoun:2023zbc} with a $6\,$MeV error, viz.
\begin{align}\label{eq:physval_etac}
m_{\eta_{\rm c}} = 2978(6)\,{\rm MeV}\,.
\end{align}
We are confident that $6\,$MeV reasonably quantifies the systematic
uncertainty on our $m_{\eta_{\rm c}}$ input, which is dominantly due to
ignoring the impact of charm annihilation to gluons.
This is supported by ref.~\cite{Hatton:2020qhk}, where the latter was
estimated to be $7.3(1.2)\,$MeV, and is also in line with calculations
that include the disconnected diagrams~\cite{Bali:2011rd,Levkova:2010ft,deForcrand:2004ia}.
The value~(\ref{eq:physval_etac}) is also among the physical inputs entering
the $N_{\rm f}=2+1$ mixed-action study~\cite{Bussone:2023kag} with maximally
twisted Wilson valence fermions.

Alternatively, one can tune the heavy-strange meson mass to the $\rm D_{\rm s}$
mass.
As explained in ref.~\cite{Bali:2017pdv}, correcting the experimental
result for the absence of electromagnetic effects that were estimated in
ref.~\cite{Goity:2007fu}, $m_{\rm D_{\rm s}}$ assumes the value
\begin{align}\label{eq:physval_Ds}
m_{\rm D_{\rm s}} = 1966.0(4)\,{\rm MeV}\,.
\end{align}
This mass is approximately constant across the ensembles
along the $\widehat{m}_{\rm s} \approx \text{const.}$ trajectory,
but varies along the $\Tr M=\text{const.}$ and symmetric lines.

A third possibility is to match the flavour average of heavy-light and
heavy-strange meson masses with the flavour averaged $\rm D$ meson mass.
Taking $m_{\rm D}= 1866.1(2)$, again from ref.~\cite{Bali:2017pdv} in the
isospin-symmetric limit of QCD and upon subtracting the electromagnetic mass
contributions, then yields as physical input for the flavour average:
\begin{align}\label{eq:physval_avgDDs}
m_{\bar{\rm D}} =
\frac{2}{3} m_{\rm D} + \frac{1}{3} m_{\rm D_{\rm s}} = 1899.4(3)\,{\rm MeV}\,.
\end{align}
This combination is our preferred choice, as it remains almost constant
along the $\Tr M=\text{const.}$ trajectory, i.e.\ at constant physical
flavour average of pion and kaon masses
(denoted as $\bar{\mathbb{M}}^2_{\rm phys}$ in section~\ref{s:setup}),
for which we have the most ensembles.
Nevertheless, we also fix the physical charm quark mass using the
$\eta_{\rm c}$ and ${\rm D_s}$ masses to check the sensitivity of our results
to the matching procedure.

Lastly, for the sake of completeness, it should be mentioned that at a
preliminary stage of our work \cite{Collins:2017iud,Collins:2017rhi},
we had also explored the spin-flavour averaged meson mass to fix the
charm quark mass: its construction as a suitable linear combination of
heavy-light and heavy-strange pseudoscalar and vector meson masses is
guided by heavy-quark symmetry, and its use was also tested in ref.~\cite{Heitger:2021apz}.
However, since the vector meson correlators are considerably noisier than
the pseudoscalar ones (even if they allow extracting the ground state
mass reliably), the statistical errors of the vector meson masses are
still relatively large.
Consequently, we do not observe any gain in overall statistical precision
when including the spin-flavour averaged meson mass in the combined
chiral-continuum fits and hence disregard it from our final analysis.

\subsection{The loss function\label{s:loss}}
To describe the dependence on the light, strange and heavy quark masses and on the lattice spacing, we use the independent variables
$t_0$, $m_\pi$, $m_\mathrm{K}$, $m_{\bar{\mathrm{D}}^1}$,
$m_{\bar{\mathrm{D}}^2}$ in the fits (or $\{m_{\rm D_{\rm s}^1}, m_{\rm D_{\rm s}^2}\}$ or $\{m_{\eta_{\rm c}^1}, m_{\eta_{\rm c}^2}\}$
in the alternative schemes). The dependent variables for
the simultaneous fit of heavy-light and heavy-strange decay constants for two
values of the heavy quark mass per ensemble are $f_{\mathrm{D_s^1}}$,
$f_{\mathrm{D_s^2}}$, $f_{\mathrm{D^1}}$ and $f_{\mathrm{D^2}}$.
We use the method described in appendix F.4 of ref.~\cite{RQCD:2022xux} to promote
independent variables to fit parameters in order to take their uncertainty and
correlation with the dependent variables into account.
For a fully correlated fit we have to compute a $9\times 9$
covariance matrix per ensemble, which reduces to $6\times 6$ on 
ensembles with SU(3) symmetry. As different Monte Carlo chains are not
correlated with each other, the full covariance matrix of the 
fit is block diagonal.

In practice, we fit the bare matrix elements of the $\mathrm{O}(a)$ improved axial vector current
defined in eq.~\eqref{e:decayc_bare} with a fit function which describes the renormalised $\mathrm{O}(a)$ improved decay constants (outlined in the next section) divided by the factor (c.f.~\cref{eq:RenImp,vectorWardIdentity})
\begin{align}
Z_{\rm A}\left(1+b_{\rm A} a m_{q\rm c}\right) =
Z_\mathrm{A}\bigg(1+b_\mathrm{A}\left(\frac{1}{4\kappa_q}+\frac{1}{4\kappa_{\rm c}}-\frac{1}{2\kappa_\mathrm{crit}}\right)\bigg)\,.
\end{align}
This approach enables us to take the uncertainties in the
renormalisation and improvement factors and the critical hopping
parameter~(which all only depend on the gauge coupling) consistently
into account. Treating these quantities as
external input, we introduce Bayesian priors and widths
(3 priors per $\beta$, with 18 priors in total) which
are set to the central values and uncertainties, respectively, listed
in \cref{tab:inputparam}.
Following ref.~\cite{RQCD:2022xux}, the full loss function that is minimised in the
fit corresponds to
\begin{align}
\chi^2 = \sum_{e=1}^{n_e}  \sum_{i,i^\prime=1}^{N}\delta f_e^i (C_e^{-1})^{ii^\prime} \delta f_e^{i^\prime}+\sum_{j=1}^{18}\left(\frac{\delta p_j}{\Delta p_j}\right)^2
\end{align}
where $n_e$ is the number of ensembles, $N=9$ or 6 is the number of observables for each ensemble, $\delta f_e$ is the difference of the data and our model, $C_e$ is the covariance matrix for ensemble $e$, $\delta p_j$ is the difference of the prior and the corresponding fit parameter and $\Delta p_j$ is the width of the prior.
We remark that the prior fit parameters, obtained from the minimisation, agree within errors with the prior values. We make use of the Cholesky-decomposition of the covariance matrix
\begin{align}
(C_e^{-1})^{ii^\prime}=(L_e)^{ik}(L_e^T)^{ki^\prime}\,,
\end{align}
to rewrite the $n_e\times N$ residuals as
\begin{align}
r_e^k=\delta f_e^i(L_e)^{ik}\,,
\end{align}
which leads to the final form
\begin{align}
\label{e:chisq}
\chi^2=\sum_{e=1}^{n_e}  \sum_{k=1}^{N}r_e^k(r_e^T)^k+\sum_{j=1}^{18}\left(\frac{\delta p_j}{\Delta p_j}\right)^2\,.
\end{align}
The loss function is minimised using the Levenberg-Marquardt algorithm in the 
\texttt{scipy} package \cite{2020SciPy-NMeth, 10.1007/BFb0067700}.

\subsection{The fit model \label{s:fitmodel}}
Having defined the loss function, we now introduce the models used to describe our data. 
We simultaneously fit to the dimensionless combinations $\sqrt{8t_0}f_{\rm D}$ and
$\sqrt{8t_0}f_{\rm D_s}$, where $t_0$ is the gradient flow scale. The
light, strange and charm quark dependence of the decay constants is
expressed in terms of the basis
\begin{align}
\bar{M}^2 \coloneqq \frac{1}{3}\big(2m_\mathrm{K}^2 + m_\pi^2\big)\propto (2m_{\rm l} + m_{\rm s})\,,\quad &\delta M^2 \coloneqq 2\big(m_\mathrm{K}^2-m_\pi^2\big) \propto (m_{\rm s} - m_{\rm l})\,,\\
\bar{M}_{\rm H} \coloneqq {} m_{\bar{\rm D}} =  \frac{2}{3} m_{\rm D}&
+ \frac{1}{3} m_{\rm D_{\rm s}} \propto m_{\rm c}\,,
\end{align}
and, similarly, we form dimensionless combinations to give,
\begin{align}
\Mbarsq \coloneqq 8t_0\bar{M}^2\,,\quad 
\dMsq \coloneqq 8t_0 \delta M^2\,,\quad 
\MbarH \coloneqq \sqrt{8t_0} \bar{M}_{\rm H}\,.
\end{align}

To begin with, we consider the light and strange quark mass dependence of the decay
constants in the continuum limit at fixed charm quark
mass. The leading terms in our ansätze
are inspired by next-to-leading order (NLO) SU(3) heavy meson chiral perturbation theory (HM$\chi$PT) \cite{Goity:1992tp} and read
\begin{align}
\sqrt{8t_0}f_{\rm D_{\mathrm{s}}}
&=
f_0 
+ c_1 \, \Mbarsq 
+ \frac{2}{3}c_2 \, \dMsq 
+ c_3 \, (4\mu_\mathrm{K}+\frac{4}{3}\mu_\eta)\,, \label{eq_fD_cont} \\
\sqrt{8t_0}f_{\rm D}
&=
f_0 
+ c_1 \, \Mbarsq 
- \frac{1}{3}c_2 \, \dMsq 
+ c_3 \, (3\mu_\pi+2\mu_\mathrm{K}+\frac{1}{3}\mu_\eta)\,, \label{eq_fDs_cont}
\end{align}
where the chiral logarithms are defined as
\begin{align}
\mu_X = 8t_0 m_X^2\log(8t_0m_X^2)
\end{align}
with $X\in\{\pi, {\rm K},\eta\}$, and the mass of the $\eta$ meson is given by the Gell-Mann-Okubo relation
\begin{align}
m_\eta^2 \approx \frac{4}{3}m_\mathrm{K}^2 - \frac{1}{3}m_\pi^2 =\bar{M}^2+\frac{1}{3}\delta M^2\,.
\end{align}
Note that SU(3) symmetry constrains the coefficients of the expansion, with $f_{\rm D}=f_{\rm D_s}$ when $\dMsq=0$, and, to this order, only four low energy constants are needed to parameterise both decay constants.
If we assume NNLO SU(3) ChPT~\cite{Bar:2013ora} for the dependence of $t_0$ on the quark masses, the rescaling of $f_{\rm D_{(s)}}$ with $\sqrt{8t_0}$ does not introduce any additional terms in~\cref{eq_fD_cont,eq_fDs_cont}.

Turning to the heavy quark mass dependence, we only have to describe
this dependence in a small region around the physical charm quark
mass. We consider terms proportional to $\MbarH$, $\dMsq\MbarH$,
$\Mbarsq \MbarH$ and $\MbarH^2$, among others~(see below). Such terms
arise when expanding in the difference between $\MbarH$ and its value
at the physical point.\footnote{An alternative approach, that is
consistent with the heavy quark limit, would be to form the
combinations $\sqrt{\vphantom{b}m_{\rm D_{(s)}}}f_{\rm D_{(s)}}$ and to expand in
powers of $1/m_{\rm D_{(s)}}$. 
Given that the (global) interpolation to the physical charm quark mass is tightly constrained
by the large number data points~(see section~\ref{s:illust}), we expect this approach to yield very consistent results.}

We use the gradient flow scale determined on each ensemble, $t_0/a^2$, to parameterise the lattice spacing dependence,
\begin{align}
\mathbb{a}^2 \coloneqq \frac{a^2}{8t_0} = \frac{a_\text{chiral}^2(g_0^2)}{8t_{0,\text{chiral}}} + \mathrm{O}(a^2 \Mbarsq)\,,
\end{align}
where $t_{0,\text{chiral}}$ is the value of $t_0$ in the chiral limit. The value of the lattice spacing in the chiral limit at $g_0^2$ is labelled as $a_\text{chiral} \equiv \lim_{M\to 0} a(g_0^2, a \Tr M)$. Note that, assuming ChPT to hold, the leading order correction (starting at $a^2 \Mbarsq$) originates from the correction to $t_0$ at NNLO SU(3) ChPT. As we are simulating at fixed $g_0^2$ (not fixed $\tilde g_0^2$), discretisation effects enter as (small) order $a^3 \Mbarsq$ corrections, and order $a^4 \Mbarsq$ and $a^4  \dMsq$ effects arise due to $\mathrm{O}(a^2)$ corrections on $t_0/a^2$. Each of the above terms is taken into account in our fit models as part of the description of the discretisation effects.

Our ansatz for the lattice spacing effects is guided by Symanzik
effective theory~\cite{Symanzik:1983dc, Symanzik:1983gh}. 
After full $\mathrm{O}(a)$ improvement of the action and observables, 
cutoff effects start at order $a^2$. Due to the
relatively large mass of the charm quark, we expect to resolve mass-dependent
cutoff effects. As outlined in refs.~\cite{Husung:2019ytz, Husung:2021mfl}, quantum corrections 
to Symanzik effective theory may introduce discretisation effects of order
$a^2 [\bar{g}^2(1/a)]^{\hat{\gamma}_i}$, where $\bar{g}^2(1/a)$ is the renormalised coupling, leading to a logarithmic modification of the cutoff effects. In refs.~\cite{Husung:2021mfl, Husung:Lattice2023} it has been found that the smallest anomalous dimension $\hat{\gamma}_i$ for pseudoscalar and 
axial vector currents is 0 for our action  and there is no dangerous slowing down of the
continuum extrapolation. We assume that contributions from terms with 
$\hat{\gamma}_i > 0$ are subleading and consider terms proportional to
$\mathbb{a}^2$, $\mathbb{a}^2\Mbarsq$, $\mathbb{a}^2\dMsq$, $\mathbb{a}^2\MbarH$ and higher powers in $\mathbb{a}$,  as detailed below.

We construct global fit models by combining the individual terms for the
dependence on the light, strange and heavy quark masses and on the inverse cutoff
additively, such that the models are linear in the fit parameters. 
By investigating the fit quality, we compose a model with a minimal number of
parameters that is able to describe our data reasonably well,
\begin{align}
\label{e:minimal_ansatz}
\sqrt{8t_0}f_{\rm D_{\mathrm{s}}}(m_\pi, m_\mathrm{K}, m_{\rm \bar{D}}, a)
=&{}
f_0 
+ c_1\, \Mbarsq 
+\frac{2}{3}c_2\, \dMsq 
+ c_3\, {(4\mu_\mathrm{K}+\tfrac{4}{3}\mu_\eta)\phantom{\frac{1}{1}\mu_\mathrm{\pi}+}\,}
+c_4\, \MbarH\\ \nonumber
&{}
+ c_5\, \MbarH^2
+ c_6\, \dMsq \MbarH
+ c_8\, \Mbarsq \MbarH
+ c_9\, \mathbb{a}^2 
+ c_{10}\, \mathbb{a}^2\MbarH
\,,
\\\nonumber
\sqrt{8t_0}f_{\rm D}(m_\pi, m_\mathrm{K}, m_{\rm \bar{D}}, a)
=&{}
f_0 
+ c_1\, \Mbarsq 
-\frac{1}{3}c_2\,  \dMsq 
+ c_3\, (3\mu_\pi+2\mu_\mathrm{K}+\tfrac{1}{3}\mu_\eta)
+ c_4\, \MbarH\\\nonumber
&{}
+ c_5\, \MbarH^2
+ c_7\, \dMsq \MbarH
+ c_8\, \Mbarsq \MbarH
+ c_9\, \mathbb{a}^2 
+ c_{10}\, \mathbb{a}^2\MbarH
\,.
\end{align}
Only eleven parameters are needed to describe the 160 naive degrees of
freedom, which are effectively reduced due to the correlation in the
data. The fit quality for this fully correlated fit is $\chi^2/{\rm
  d.o.f}=1.08$. Omitting terms in eq.~\eqref{e:minimal_ansatz},
e.g.\ setting any of $c_{5-8,10}$ to zero, leads to a significant
increase in the $\chi^2$.
Note that the above expressions are consistent with SU(3) constraints
and there is only one fit parameter ($c_6$ respectively $c_7$) that is not shared by the
ansätze for $f_{\rm D}$ and $f_{\rm D_{\mathrm{s}}}$.
We find that, in addition to the linear term, we have to incorporate a
term that is quadratic in $\MbarH$ which could not have been resolved when
fixing the charm quark mass ensemble by ensemble prior to the extrapolation.
Furthermore, mixed terms involving the heavy quark mass and the sea quark mass proxies
are necessary to achieve a good description of the data.
In terms of lattice spacing effects, mass-independent contributions and those dependent on
the heavy quark are dominant.

To explore the parameter space of the extrapolation and to test for higher order
effects, we build a variety of models, which extend
\cref{e:minimal_ansatz}. We add up to four terms out of the following list
of higher order terms in the quark masses,
\begin{align}
	\Mbarsq \MbarH^{2}\,, \enspace
	\Mbarsq \dMsq\,, \enspace
	(\dMsq)^{2}\,, \enspace
	\dMsq \MbarH^{2}\,, \enspace
	(\dMsq)^{2}\MbarH\,,
\end{align}
and up to three terms out of the following lists of terms describing lattice artifacts,
\begin{align}
	\mathbb{a}^2 \Mbarsq\,, \enspace
	\mathbb{a}^2 \dMsq\,, \enspace
	\mathbb{a}^3\,, \enspace
	\mathbb{a}^3 \Mbarsq\,, \enspace
	\mathbb{a}^3 \dMsq\,, \enspace
	\mathbb{a}^3 \MbarH\,, \enspace
	\mathbb{a}^4\,, \enspace
	\mathbb{a}^4 \Mbarsq\,, \enspace
	\mathbb{a}^4 \dMsq\,, \enspace
	\mathbb{a}^4 \MbarH\,.
\end{align}
We exclude models that mix $a^3$ and $a^4$ cutoff effects and models with
more than 16 parameters. 
In total this amounts to $K=482$ models. 
The worst fit quality found in this set has a fully correlated
$\chi^2/\mathrm{d.o.f.}=1.09$. 
The best fit quality of the models under consideration is found to be
$\chi^2/\mathrm{d.o.f.}=0.92$ for the model,
\begin{align}
\label{e:best_ansatz}
\sqrt{8t_0}f_{\rm D_{\mathrm{s}}}(m_\pi, m_\mathrm{K}, m_{\rm \bar{D}}, a)
=&{}
f_0 
+ c_1\, \Mbarsq 
+\frac{2}{3}c_2\, \dMsq 
+ c_3\, {(4\mu_\mathrm{K}+\tfrac{4}{3}\mu_\eta)\phantom{\frac{1}{1}\mu_\mathrm{\pi}+}\,}
+c_4\, \MbarH\\ \nonumber
&{}
+ c_5\, \MbarH^2
+ c_6\, \dMsq \MbarH
+ c_8\, \Mbarsq \MbarH
+ c_9\, \mathbb{a}^2 
+ c_{10}\, \mathbb{a}^2\MbarH
\\\nonumber &{}
+c_{11}\, \dMsq \MbarH^{2}
+c_{13}\, \mathbb{a}^3
+c_{14}\, \mathbb{a}^3 \dMsq
\,,
\\\nonumber
\sqrt{8t_0}f_{\rm D}(m_\pi, m_\mathrm{K}, m_{\rm \bar{D}}, a)
=&{}
f_0 
+ c_1\, \Mbarsq 
-\frac{1}{3}c_2\,  \dMsq 
+ c_3\, (3\mu_\pi+2\mu_\mathrm{K}+\tfrac{1}{3}\mu_\eta)
+ c_4\, \MbarH\\\nonumber
&{}
+ c_5\, \MbarH^2
+ c_7\, \dMsq \MbarH
+ c_8\, \Mbarsq \MbarH
+ c_9\, \mathbb{a}^2 
+ c_{10}\, \mathbb{a}^2\MbarH
\\\nonumber &{}
+c_{12}\, \dMsq \MbarH^{2}
+c_{13}\, \mathbb{a}^3
+c_{14}\, \mathbb{a}^3 \dMsq
\,.
\end{align}
Especially the inclusion of higher order cutoff effects, 
such as the terms multiplying $c_{13}$ and $c_{14}$ in \cref{e:best_ansatz},
is found to improve the fit quality with respect to the base fit.

\subsection{Model averages \label{s:model}}
For our final result, we take into account all the information that we derive from
the set of $K$ fits defined above, by performing a weighted average over 
the results. The weight that is assigned to a model $k$
is based on the Akaike information criterion (AIC) \cite{1100705, Akaike1998}, 
\begin{align}
	\label{e:AIC}
	{\rm AIC}_k = \chi^2_k + 2M_k\,,
\end{align}
where $M_k$ is the number of model parameters and the loss function $\chi^2$ 
is defined in \cref{e:chisq}. 
The AIC introduces a penalty 
for each additional fit parameter, such that fits with less 
parameters are preferred over fits with more parameters but similar $\chi^2$.
The weight for each model reads
\begin{align}
	\label{e:AIC_weight}
	w_k^\mathrm{AIC}=N\exp\left(-\frac{1}{2}{\rm AIC}_k\right)\,,
\end{align}
where the normalisation $N$ is chosen such that $\sum_{k=1}^K w_k^\mathrm{AIC}=1$.
The central value for an observable $\mathcal{O}$ 
is computed from a weighted average over all models,
\begin{align}
	\label{e:mav_value}
	\mathcal{O} = \sum_{k=1}^{K} w_k^\mathrm{AIC} \mathcal{O}_k\,,
\end{align}
where $\mathcal{O}_k$ is the result for the observable based on model $k$. 
The statistical error is also obtained from this weighted average via standard
error propagation for derived observables.
Following ref.~\cite{Jay:2020jkz}, we estimate the systematic uncertainty, stemming
from the variation in the model space, from the width of the distribution of
the results in the model average,
\begin{align}
	\label{e:mav_syst}
	\sigma_\mathrm{sys}^2
	=
	\sum_{k=1}^{K} w_k^\mathrm{AIC} \mathcal{O}_k^2 - \left( \sum_{k=1}^{K} w_k^\mathrm{AIC} \mathcal{O}_k \right)^2\,.
\end{align}

The fit with the largest model weight of $0.025$ is also the fit with
the smallest $\chi^2/\mathrm{d.o.f.}$, i.e.\ the one of \cref{e:best_ansatz}.
The fit with the smallest contribution to the average has a weight of
$2\cdot10^{-6}$.  The model average does not include cuts on the data, e.g.\ 
excluding the data points for the largest value of $\Mbarsq$ or
$\mathbb{a}$. This is because the base model in
\cref{e:minimal_ansatz}, without higher order terms in lattice spacing 
or quark masses, adequately represents the entire dataset.
However, we explicitly test that fitting the base model remains
consistent when applying cuts in the lattice spacing, the pion mass, 
the sea quark parameter $\Mbarsq$ or the heavy quark parameter $\MbarH$. 
Among this set of cuts, the most pronounced impact stems from cuts in the
lattice spacing. The exclusion of ensembles at the
coarsest lattice spacings leads to a small downward shift of the decay
constants (and a slight upward shift of their ratio). A similar effect is
observed when including terms that allow for $\mathrm{O}(a^3)$ or
$\mathrm{O}(a^4)$ cutoff effects when fitting the full data set, as it is
the case in \cref{e:best_ansatz}.

\subsection{Illustration of the quark mass and cutoff dependence\label{s:illust}}
\begin{figure}[t]
	\centering
	\includegraphics[width=0.9\textwidth]{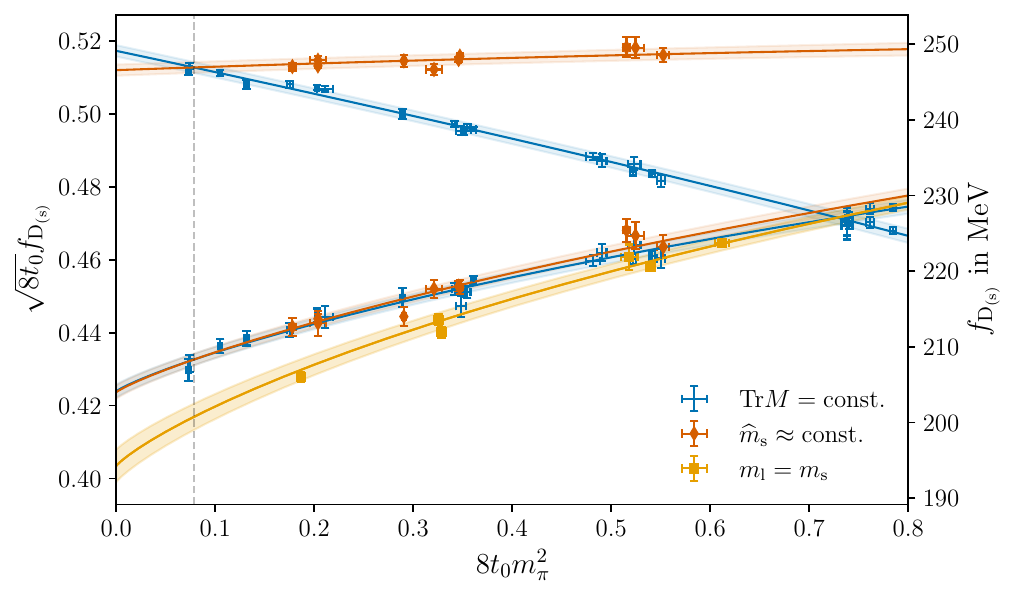}
	\caption{Dependence of $f_{\rm D}$ and $f_{\rm D_s}$ on the
          pion mass squared for the fit with the largest
          AIC weight~(eq.~\eqref{e:best_ansatz}), with all quantities in dimensionless combinations with
          the gradient flow scale $t_0$. The data points  are
          corrected for effects due to the cutoff and mistuning of
          the light and strange quark mass and shifted to the physical charm quark mass, see the
          text. The results lie on three trajectories, for which $\Tr
          M = \text{const.}$, $\widehat{m}_{\rm s} \approx
          \text{const.}$ and $m_{\rm l}=m_{\rm s}$, as described in
          section~\ref{s:setup} and displayed in
          figure~\ref{fig:CLS}. The central value and error band of the fit
          in the continuum limit, for the physical charm quark mass, is projected onto these trajectories.
          The vertical dashed line indicates
          the physical point.}
	\label{fig:pion_mass_dependence}
\end{figure}

\begin{figure}[t]
	\centering
	\includegraphics[width=0.9\textwidth]{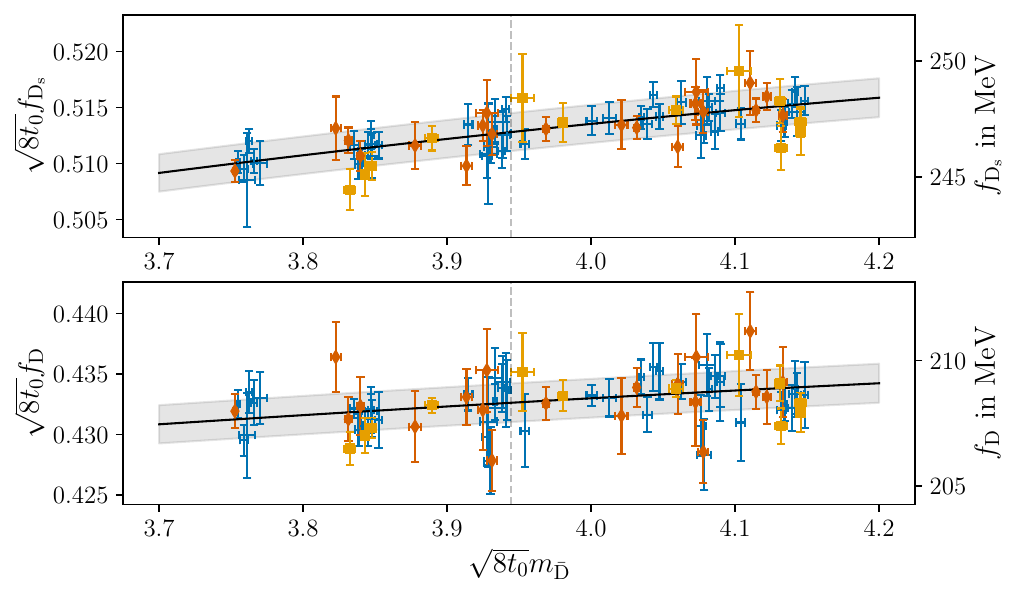}
	\caption{Dependence of $f_{\rm D}$ and $f_{\rm D_s}$ on the
          flavour average D meson mass for the fit with the largest AIC
          weight, as shown in
          figure~\ref{fig:pion_mass_dependence}. The dashed line
          indicates the physical value of $\sqrt{8t_0}m_{\bar{\rm
              D}}$.  The data points are shifted using the fit to
          the physical pion and kaon masses and lattice spacing
          effects are removed.
          The central value and error for the
          corresponding $m_{\rm \bar{D}}$ dependence of the fit are
          shown as the black line and grey band, respectively. }
	\label{fig:D_mass_dependence}
\end{figure}
We use the fit with the largest model
weight~(eq.~\eqref{e:best_ansatz}) to illustrate the dependence of
$f_{\rm D}$ and $f_{\rm D_{\rm s}}$ on the quark masses and the
cutoff. In figure~\ref{fig:pion_mass_dependence}, we display the light
quark mass dependence, where the fit is used to correct the results
for lattice spacing effects, mistuning of the light and strange quark
masses~(see the discussion in section~\ref{s:setup}) and to shift to
the physical charm quark mass. In terms of the mistuning, for a given
pion mass, the strange quark mass is fixed implicitly by the chiral
trajectories: the data on the $\Tr M=\text{const.}$ and
$\widehat{m}_{\rm s} \approx \text{const.}$ trajectories are shifted to
correspond to the kaon mass which gives the physical value of
$\bar{M}^2=(2m_{\rm K}^2+m_\pi^2)/3\propto 2m_{\rm l}+m_{\rm s}$ and $2m_{\rm
  K}^2 - m_\pi^2 \propto m_{\rm s}$, respectively.  At the physical
point, the continuum fit curves projected onto these two
trajectories~(coloured blue and orange, respectively) have to coincide
by definition, tightly constraining the fit.  The fit curves clearly
show that the strange quark effects on $f_{\rm D}$ are small, while
$f_{\rm D_{\rm s}}$ is largely insensitive to the light quark mass,
when keeping the strange quark mass at its physical value.

Along the symmetric line~(yellow curve), which approaches the SU(3)
chiral limit when lowering the pion mass, the $f_{\rm D}$ and $f_{\rm
  D_s}$ decay constants are equal.  Including the data on this
trajectory helps to constrain the dependence on the parameter
$\Mbarsq$.  Flavour breaking effects are observed for the results on
the $\Tr M=\text{const.}$ trajectory, starting from
the SU(3) symmetric point and decreasing the pion mass towards
the physical point. The curvature due to the chiral logarithms can
be mapped out thanks to the two ensembles at physical pion mass and
several further ensembles with $m_\pi < 200\,$MeV.

The heavy quark mass dependence of the decay constants, parameterised
by $\MbarH$ in eq.~\eqref{e:best_ansatz}, is presented in figure
\ref{fig:D_mass_dependence}. Note that the data points are corrected
for all cutoff effects~(including those arising from the $\mathbb{a}^2\MbarH$ terms) and
shifted to correspond to the physical light and strange quark
masses. Two charm quark masses are realised per ensemble and, overall,
the results bracket the physical value of the flavour average D meson
mass, indicated by the dashed line.  By performing a global fit of the
heavy quark mass dependence, we are able to resolve terms quadratic in
$\MbarH$. However, as seen in the figure, these contributions are rather minor.

\begin{figure}[t]
\centering
\includegraphics[width=0.9\textwidth]{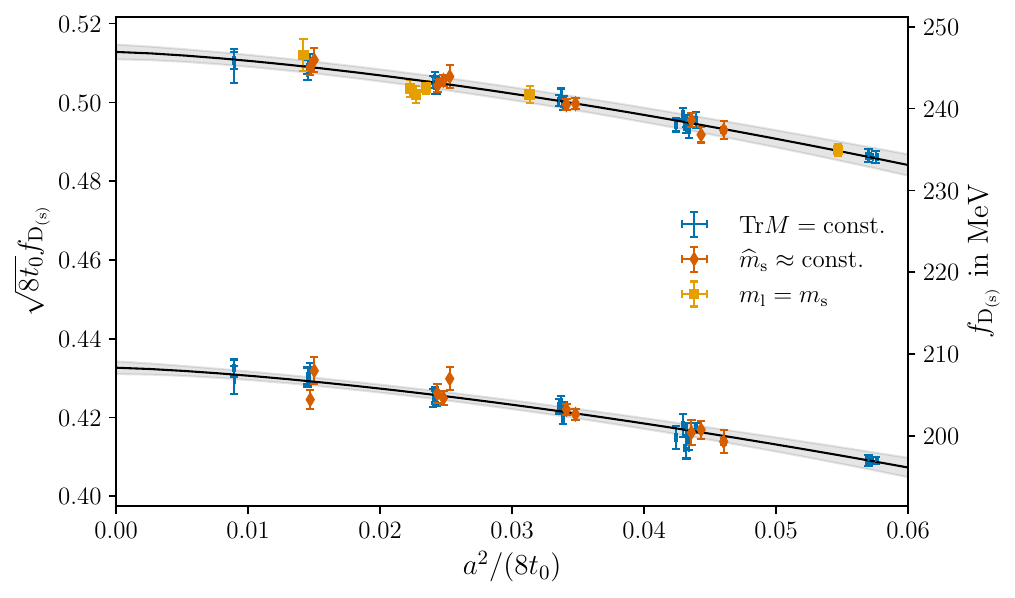}
\caption{Dependence of $f_{\rm D}$ and $f_{\rm D_s}$ on the lattice
  spacing squared for the fit with the largest AIC weight, as shown in
  figure~\ref{fig:pion_mass_dependence}. The data points are shifted
  using the fit to the physical pion, kaon and $\bar{\rm D}$ meson masses.
  The
  central value and error for the corresponding lattice spacing dependence of the
  fit are shown as the black line and grey band, respectively. }
\label{fig:a2_dependence}
\end{figure}

\begin{figure}[t]
	\centering
	\includegraphics[width=0.9\textwidth]{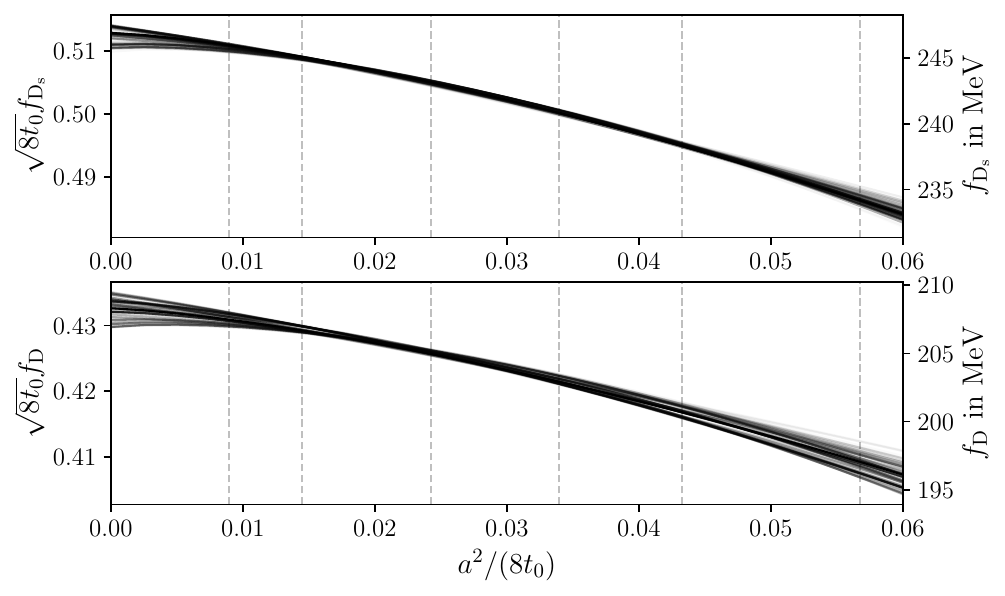}
	\caption{Illustration of the lattice spacing dependence of all fits considered. Each line corresponds to the central value of one model that enters the model average. The opacity of a line indicates the relative model weight, with a black line corresponding to the highest weight. The dashed lines show the six values of the lattice spacing where the fit is constrained by data.
	} 
	\label{fig:cont_spread}
\end{figure}

\Cref{fig:a2_dependence} shows the projection of the best fit onto
the cutoff dependence for physical quark masses. Discretisation
effects are a significant source of systematics for observables
involving charm quarks, with $am_{\rm c} \approx 0.6$ for our coarsest
lattice spacing. However, by utilising high statistics data at six
lattice spacings ranging from $a \approx 0.10\,$fm down to
$a=0.039\,$fm~($a^2$ varies by more than a factor of 6), we are able to
clearly resolve the lattice spacing dependence, including both
$\mathbb{a}^2$ and $\mathbb{a}^3$ terms. With full non-perturbative
$\mathrm{O}(a)$ improvement, the size of these effects is fairly moderate with
a 5\% difference between the decay constants at the coarsest lattice
spacing and in the continuum limit. When varying the fit model, we
found that fits including $\mathbb{a}^3$ or $\mathbb{a}^4$ terms were favoured by the
data.  This is illustrated in figure~\ref{fig:cont_spread} which shows
the lattice spacing dependence for all fits considered.
The variation of the ansatz to describe the cutoff effects, as illustrated
in the figure, is the main contribution to our final systematic uncertainty. 

\section{Results \label{s:results}}
Employing the isospin-corrected experimental pion, kaon and flavour
average $\rm D$ meson masses given in section~\ref{s:scheme} and the
physical value of the flow scale scale
of ref.~\cite{RQCD:2022xux}, we extract results for
the decay constants at the physical point~(in the continuum limit) in
physical units for each of the 482 fits considered.  For the ratio
$f_{\mathrm{D_s}}/f_\mathrm{D}$, we simply divide the results at the
physical point for the individual decay constants. Alternatively, one
can perform a simultaneous continuum, chiral extrapolation of the
ratio directly. While only a reduced number of terms would need to be
included in the model for such a fit (many terms cancel when taking
the ratio of the two fit forms, for instance, in \cref{e:best_ansatz}),
the data set entering the fit would be much reduced. In particular,
results from ensembles with SU(3) symmetry would not be included. As
our simultaneous fits to $f_{\rm D}$ and $f_{\rm D_s}$ take all
correlations into account, we would not expected to obtain more
precise results by fitting to the ratio.
Following the model averaging procedure outlined in section~\ref{s:model},
our final results at the physical point of $N_{\rm f}=2+1$ flavour QCD read
\begin{align}\label{e:final_results}
f_{\mathrm{D_s}} &= 246.8(0.64)_{\rm stat}(0.61)_{\rm sys}(0.95)_{\rm scale}[1.3]\,\mathrm{MeV}\,, \nonumber\\
f_\mathrm{D} &= 208.4(0.67)_{\rm stat}(0.75)_{\rm sys}(1.11)_{\rm scale}[1.5]\,\mathrm{MeV}\,,\\
f_{\mathrm{D_s}}/f_\mathrm{D} &= 1.1842(21)_{\rm stat}(22)_{\rm sys}(19)_{\rm scale}[36]\,, \nonumber
\end{align}
where the first error is statistical, the second is due to the
systematics and the third arises from the scale setting. The
statistical error includes the uncertainties due to the
renormalisation and improvement coefficients and the hadronic scheme,
while the systematic error quantifies the uncertainty stemming from the
model variation for continuum and quark mass extrapolations or interpolations.
The total
uncertainty obtained by adding the individual errors in quadrature is
given within the square brackets. We achieve a $0.5\%$, $0.7\%$ and
$0.3\%$ overall error in $f_{\mathrm{D_s}}$, $f_{\mathrm{D}}$ and
$f_{\mathrm{D_s}}/f_\mathrm{D}$, respectively. The results for the
ratio indicate that SU(3) flavour symmetry breaking effects in the
decay constants are around 20\% at the physical point.

\begin{figure}[t]
	\centering
	\includegraphics[width=0.9\textwidth]{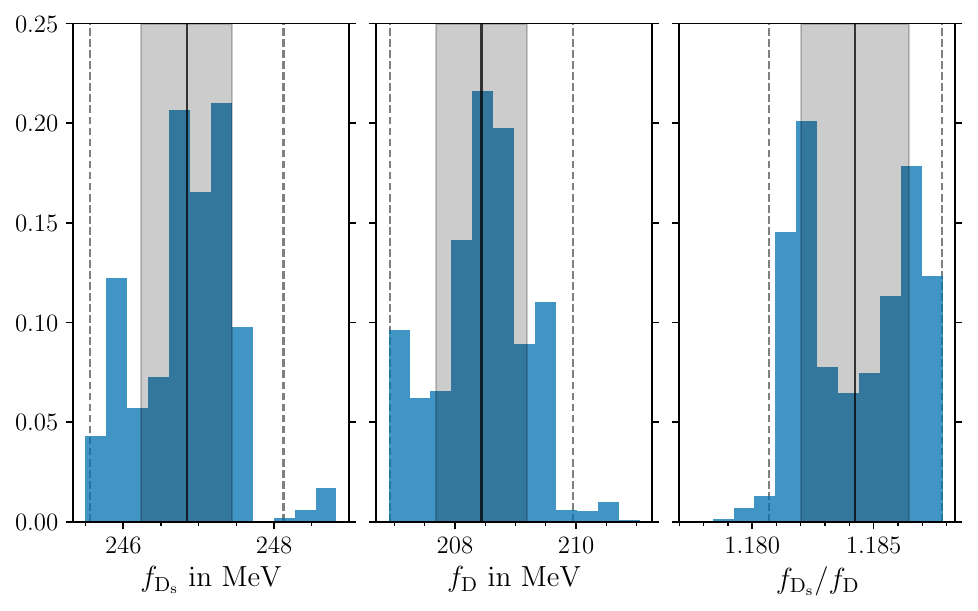}
	\caption{AIC weighted histograms of the central values of
          $f_{\rm D_s}$, $f_{\rm D}$ and their ratio obtained from the 482 fits
          that enter our model averaging procedure. The solid lines
          indicates the central values of the final results, the grey bands display the
          corresponding systematic errors and the dotted lines show the
          total uncertainties.}
	\label{fig:histogram}
\end{figure}

We illustrate the variation in the results 
across the 482 fits in figure~\ref{fig:histogram}. This shows the
distributions of the central values, where each fit contributes to the
histograms according to its AIC weight.
For both the individual decay constants and the ratio,
the spread in the values is mainly due to the variation in the
parametrisation of the discretisation effects, i.e.\ the uncertainty
arising from the continuum limit extrapolation dominates the
systematic error~(indicated by the grey band). In particular, a slight double peak structure
visible in the histogram for the ratio is due to the inclusion or
exclusion of $\mathrm{O}(a^3)$ or $\mathrm{O}(a^4)$ cutoff effects in the
model. However, this variation is well within the full uncertainty of
the final result~(shown as dashed lines).

The full error budget can be found in table
\ref{tab:error_budget}. The uncertainty due to the scale setting
dominates the total in the case of $f_{\rm D}$ and $f_{\rm D_s}$ and
contributes to the error of the ratio via the definition of the
physical point. The systematic uncertainties are of a similar size as
the statistical uncertainties from the fluctuations of the gauge
configurations. The renormalisation and improvement
coefficients~\cite{DallaBrida:2018tpn,Bali:2021qem} are determined
precisely enough that they only account for a few percent of the total
squared error. Similarly, the errors due to the physical inputs that define
the physical point are negligible. While the systematic uncertainties
are computed according to eq.~(\ref{e:mav_syst}), we track the
contribution of all other sources of error to the variance as part of
the error propagation in the $\Gamma$ method.

\begin{table}[t]
	\centering
	\caption{Relative contributions to the squared total errors of the final results.}
	\label{tab:error_budget}
	\begin{tabular}{lrrr}
		\toprule
		 & $f_{\mathrm{D_s}}$ & $f_\mathrm{D}$ & $f_{\mathrm{D_s}}/f_\mathrm{D}$ \\
		 \midrule
		Scale setting & 53.5\% & 55.1\% & 27.8\% \\
		Systematic error & 22.3\% & 24.8\% & 39.0\% \\
		Statistical error & & & \\
		\quad Gauge ensembles & 20.5\% & 18.3\% & 32.0\% \\
		\quad Renormalisation and improvement & 3.6\% & 1.8\% & 0.8\% \\
		\quad Physical meson masses & 0.1\% & 0.0\% & 0.4\% \\
		\bottomrule
	\end{tabular}
\end{table}

In our $N_{\rm f}=2+1$ simulation, where the charm quark is partially
quenched, we neglect strong isospin breaking and QED effects, as well
as the effect of charm quark loops in the sea. For low-energy
observables, the latter are expected to be negligible due to the
decoupling of the heavy quark \cite{Weinberg:1980wa}. However, for
heavy-light quantities, for which there is no clear separation of
scales, the omission of charm loops may lead to a small but
significant effect. Parametrically, this effect is of
$\mathrm{O}(\alpha_{\rm s} (\Lambda_{\rm QCD}/2m_{\rm c})^2)$
\cite{Nobes:2005yh,FermilabLattice:2016ipl} for the decay constants,
which amounts to about $0.5\%$.\footnote{For this and the following estimates quoted in the text,
we use result for $\Lambda_{\rm QCD}$ from
ref.~\cite{Bruno:2017gxd}, the charm quark mass as calculated in
ref.~\cite{Heitger:2021apz}, the values of the light and strange quark masses quoted in ref.~\cite{Bruno:2019vup} and the determination of $m_u-m_d$ presented in ref.~\cite{Bali:2023sdi}.}  In the ratio of the two decay constants, this
effect is further suppressed by $(m_{\rm s} - m_{\rm d}) /
\Lambda_{\rm QCD} \sim 0.4$, resulting in an
overall effect of about $0.2\%$.  In ref.~\cite{Cali:2021xwh}, the
effect of charm quark loops on charmonium decay constants has been
investigated non-perturbatively. Comparing the pseudoscalar decay constants
obtained from a pure Yang-Mills background with their counterparts
from a sea with two charm quarks, the authors found an increase at the level of
$0.5\%$.  With respect to our work, this represents a very
conservative upper bound since it quantifies the effect from two heavy
sea quarks for an observable with two heavy valence quarks.

Turning to the effects of strong isospin breaking, with light quarks
in the valence sector, the leading effects for $f_{\rm D}$ are linear
in the up and down quark mass difference and are of
$\mathrm{O}((m_{\rm u}-m_{\rm d})/\Lambda_{\rm QCD})\sim 0.6\%$.  So far, these effects
have only have been estimated on the lattice by FNAL/MILC in their
$N_{\rm f}=2+1+1$ study~\cite{Bazavov:2017lyh} by tuning to the physical up
and down quark masses. They find $f_{\rm D^+}-f_{\rm D}=0.58(7)$\,MeV
and $f_{\rm D^+}-f_{\rm D^0}=1.11(15)$\,MeV. This is consistent with
QCD sum rule estimates from Lucha et al., who find $f_{\rm D^+}-f_{\rm
  D^0}=0.97(13)$\,MeV~\cite{Lucha:2016nzv}.  For $f_{\rm D_s}$, strong isospin
breaking effects only arise due to the light sea quarks. They are quadratic in the
mass difference and are likely to be much smaller in magnitude than for $f_{\rm D}$.

Electromagnetic interactions are expected to enter at the level of
$\mathrm{O}(\alpha_{\rm QED})\sim 1\%$. As it is difficult to separate leptonic
decays including a photon in the final state from those only with a
lepton and a neutrino in experiment, both virtual and real radiative corrections to
the decay rate need to be considered. In this case, the decay rate can
no longer be factorised into a decay constant and a term involving the
relevant CKM matrix element. The experimental decay rates are adjusted
using estimates of these radiative effects~(see, for example, the PDG
review of leptonic decays of charged pseudoscalar
mesons~\cite{ParticleDataGroup:2022pth}) in order to extract the
combinations $f_{\rm D}|V_{\rm cd}|$ and $f_{\rm D_s}|V_{\rm cs}|$ in a
particular scheme.  Lattice calculations of the leading QED
corrections, including the determination of the form factors for
radiative leptonic decays, such as ${\rm D_s}\to
\ell\nu\gamma$~\cite{Desiderio:2020oej,Giusti:2023pot}, are ongoing.

We conclude that the effect of the missing charm quark loops and
isospin breaking in our calculation is likely to be around the same
size as or below our total uncertainties.  However, the absence of
these effects also impacts on the predictions for the decay constants
indirectly through the determination of the lattice scale and the
tuning of the quark masses. As the precision increases, the results
will depend on the hadronic scheme, with a number of different schemes
being employed in the literature. In order to facilitate a close
comparison with other lattice studies (which use $t_0$ to set the
scale), we give the dependence of the decay constants and their ratio on the input parameters that define our scheme
at linear order in appendix~\ref{a:scheme}. This allows the results to
be shifted to take into account small modifications of the scheme.  In
the appendix, we also collect the values of the decay constants for
the two schemes where the charm quark mass is fixed from the mass of
the $\rm D_{\rm s}$ and the $\eta_{\rm c}$ meson. The choice of scheme
has very little impact on the results.

From our results for $f_\mathrm{D}$ and $f_{\mathrm{D_s}}$, we can compute the
CKM matrix elements $|V_{\rm cd}|$ and $|V_{\rm cs}|$. The PDG 
\cite{ParticleDataGroup:2022pth} lists the products of decay constants 
and CKM matrix elements, derived from the experimental ${\rm D}^+\to\ell\nu[\gamma]$ and ${\rm D}^+_{\rm s}\to\ell\nu[\gamma]$ decay rates, as
\begin{align}\label{e:fDV}
f_{\rm D^+}|V_{\rm cd}| =  45.82(1.10)\,{\rm MeV}\,,\qquad 
f_{\rm D_{\rm s}^+}|V_{\rm cd}| = 243.5(2.7)\,{\rm MeV}\,,
\end{align}
respectively, where the uncertainty is due to the estimation of the radiative corrections
and the measured branching fractions. 
Together with our results in eq.~(\ref{e:final_results}), we obtain
\begin{align} \label{e:res_VcX}
|V_{\rm cd}| = 0.2199(15)(52)[55]\,,\qquad 
|V_{\rm cs}| = 0.987(5)(10)[13]\,,
\end{align}
where the first error is due to our lattice results
and the second arises from the combined experimental and non-lattice
theory uncertainty in eq.~(\ref{e:fDV}).  Note that we have not
included the systematic uncertainties due to the omission of
strong isospin breaking and charm loop effects
(estimated previously). These uncertainties
are small compared to the total error, which is dominated by the
combined non-lattice uncertainty. The above values are consistent
with, but slightly less precise than, those quoted by the PDG (that are derived
from the same leptonic decay
rates~\cite{ParticleDataGroup:2022pth}). The PDG utilise the FLAG
report $N_{\rm f}=2+1+1$
results~\cite{FlavourLatticeAveragingGroupFLAG:2021npn} for the decay
constants, which have smaller errors than ours, as discussed below. The
CKM matrix elements can also be determined from the decay rates for
the semi-leptonic decays ${\rm D}\to\pi\ell\nu$ and ${\rm D_s}\to {\rm K}\ell\nu$ and
lattice calculations of the corresponding form factors. At present,
the PDG value for $|V_{\rm cd}|$ obtained from the leptonic decay is more precise, while for
$|V_{\rm cs}|$ the value extracted from the semi-leptonic decay rate has smaller uncertainties.

\begin{figure}[t]
	\centering
	\includegraphics[width=0.8\textwidth]{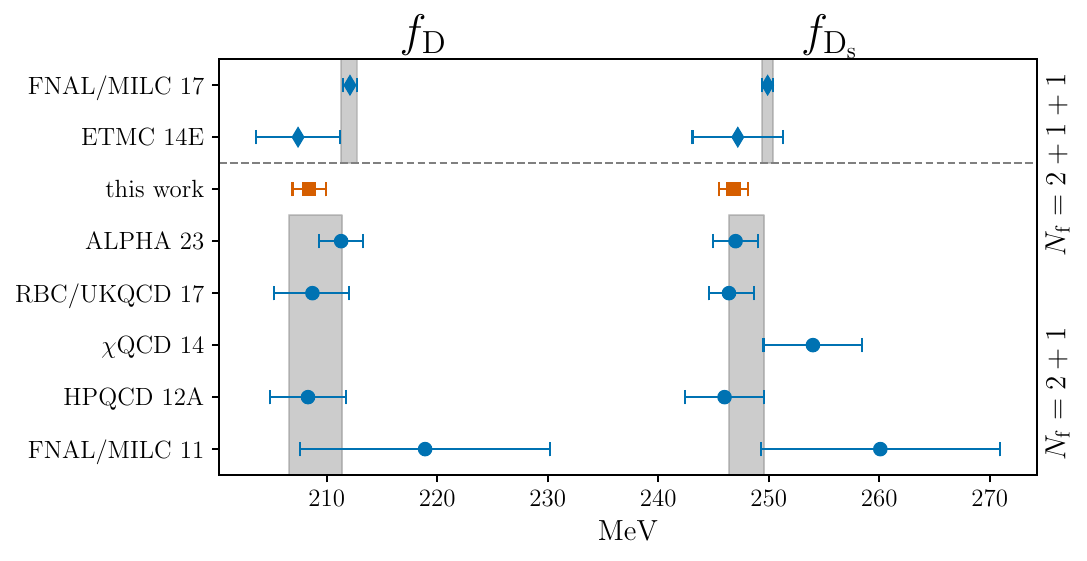}\\
	\includegraphics[width=0.8\textwidth]{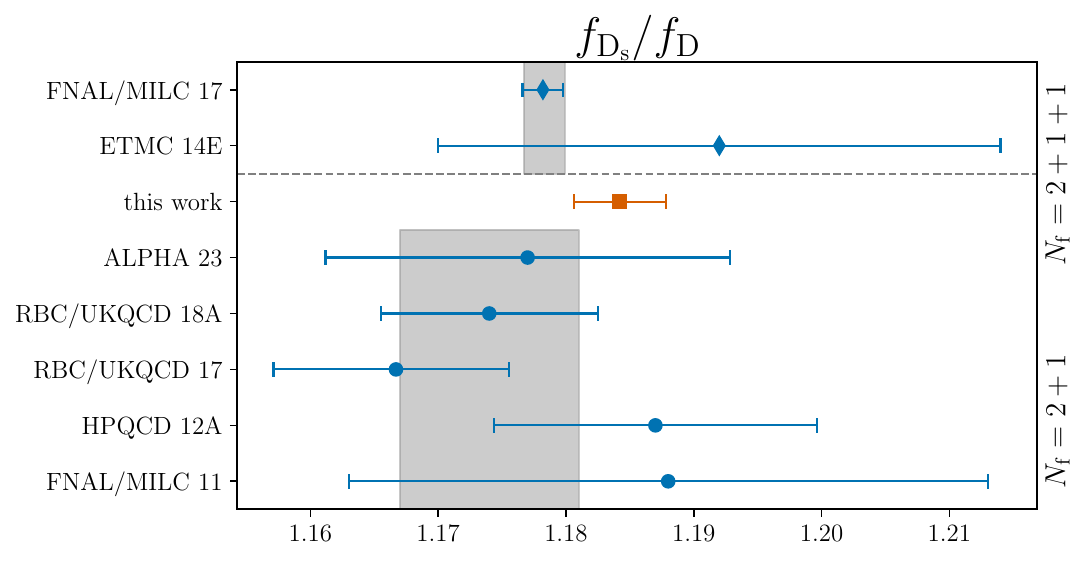}%
	\caption{
Comparison of lattice results for the decay constants $f_\mathrm{D}$ and $f_\mathrm{D_s}$~(top) and their ratio~(bottom) for $N_{\rm f}=2+1$ \cite{Davies:2010ip,FermilabLattice:2011njy,Na:2012iu,Yang:2014sea,Boyle:2017jwu,Boyle:2018knm,Bussone:2023kag} and $N_{\rm f}=2+1+1$ \cite{Carrasco:2014poa,Bazavov:2017lyh}.  Only results that fulfil all the FLAG quality criteria and that are not superseded by later works are displayed. The grey bands show the FLAG 21 averages from ref. \cite{FlavourLatticeAveragingGroupFLAG:2021npn}.
	}
	\label{fig:comparison}
\end{figure}

There is a long history of calculating $f_\mathrm{D}$ and
$f_\mathrm{D_s}$ on the lattice. In figure~\ref{fig:comparison}, we
compare our values with recent $N_{\rm f}=2+1$ and 
$N_{\rm f}=2+1+1$ determinations. Only those results that consider all
sources of systematic uncertainty in their analysis and pass 
the quality criteria of the FLAG 21 review  
\cite{FlavourLatticeAveragingGroupFLAG:2021npn}~(for the continuum limit,
chiral and finite volume extrapolations, renormalisation and the treatment of the heavy quark) are shown. We also only display the latest results for each collaboration.  Further recent
works can be found in refs.~\cite{PACS-CS:2011ngu, Chen:2020qma,
  Dimopoulos:2021qsf}. Note that the ALPHA 23 study of ref.~\cite{Bussone:2023kag}
utilises a small subset of the ensembles employed in the present analysis and we
expect some statistical correlation with our values.
They include ten ensembles with $m_\pi \geq 200\,$MeV on the $\Tr M = \text{const.}$
trajectory in a mixed action setup with maximally twisted Wilson valence 
fermions.

For the $N_{\rm f}=2+1$ theory, our results are the most precise and
represent a significant improvement on earlier studies. All works are
in reasonable agreement with each other.
FNAL/MILC~\cite{Bazavov:2017lyh} quote the smallest total
uncertainties of around $0.3(0.2)$\% for $f_{\rm D}(f_{\rm D_s})$ and
$0.1$\% for $f_{\rm D_s}/f_{\rm D}$, and their results dominate the
FLAG average for $N_{\rm f}=2+1+1$. At this level of precision, the
definition of isospin symmetric QCD has a significant impact on the
values of the decay constants.  Note that we compare to the isospin
symmetric result for $f_{\rm D}$ given in ref.~\cite{Bazavov:2017lyh}
and follow the FLAG
report~\cite{FlavourLatticeAveragingGroupFLAG:2021npn} by rescaling
the central value given for $f_{\rm D_s}/f_{\rm D^+}=1.1749(16)$ by
the ratio of the central values for $f_{\rm D^+}$ and $f_{\rm D}$ to
obtain $f_{\rm D_s}/f_{\rm D} = 1.1782(16)$.  The FNAL/MILC results
for the individual decay constants lie roughly $2\sigma$ above ours,
while the results for the ratio are slightly more consistent. These
differences are not statistically significant and, as argued above,
the absence of charm quark loop effects in our simulations is likely
to lead to a less than one sigma effect.

\section{Conclusions \label{s:conc}}
In this work, we determine the leptonic decay constants of the $\rm D$
and $\rm D_{\rm s}$ mesons in $2+1$ flavour lattice QCD with Wilson
fermions. We utilise 49 high statistics ensembles at six values of the
lattice spacing, which lie on three distinct quark mass trajectories
covering a wide range of light and strange quark masses. This enables
us to achieve an excellent description of both the cutoff effects and
the quark mass dependence down to the physical point. The charm-light
and charm-strange decay constants are fitted simultaneously, with all
correlations taken into account, and a large set of parameterisations
is considered in order to reliably quantify the systematics associated
with the continuum limit and chiral extrapolation.  Our final results
are summarised in eq.~\eqref{e:final_results} and a full error budget
is given in table~\ref{tab:error_budget}. A $0.5\%$, $0.7\%$ and
$0.3\%$ overall uncertainty is achieved for $f_{\mathrm{D_s}}$,
$f_{\mathrm{D}}$ and $f_{\mathrm{D_s}}/f_\mathrm{D}$,
respectively. These are the most precise $2+1$ flavour results
to-date.  When simulating isospin symmetric QCD, it is important to
specify the hadronic scheme used to set the scale and tune the quark
masses. In Appendix~\ref{a:scheme} we give the dependence of the
results on the hadronic input, allowing for a
close comparison with other works, where $t_0$ is used to set the
lattice scale.

Further improvement in the determination of the decay constants would
require a more precise evaluation of the scale, followed by a
reduction in both the statistical and systematic errors. The main
source of the latter error is the continuum extrapolation. The
discretisation effects are moderate but significant, with a 5\%
difference in the values of the decay constants at the coarsest
lattice spacing and those in the continuum limit. Furthermore, we
resolve cutoff effects beyond the leading order of $\mathrm{O}(a^2)$. Additional
ensembles at the finest lattice spacing of $a=0.039\,$fm with quark
masses closer to their physical values would help to further constrain the
extrapolation.  Generating even finer lattices than the ones employed
in this work is challenging, even with state-of-the-art algorithms,
due to the critical slowing down that is observed towards the
continuum limit \cite{Schaefer:2010hu}. 

The charm quark is partially quenched in our analysis.  The effect of
charm quark loops on the decay constants may not depend strongly on
the light and strange quark masses, and an estimate of the size of
this effect could be made at the SU(3) symmetric point~(corresponding
to $m_\pi=m_{\rm K}\approx 410$\,MeV), using the ensembles of this work and
the $3+1$ flavour ensembles of
ref.~\cite{Hollwieser:2020qri}. However, the renormalisation factor
for the axial vector current for the $3+1$ flavour theory, along with
the associated ${\rm O}(a)$ improvement coefficients, would need to be
computed for such comparison to be meaningful.

\acknowledgments
We thank Gunnar Bali, Tomasz Korzec, Stefan Schaefer and Rainer Sommer
for useful discussions, as well as Kevin Eckert and Stefan Hofmann for their
contributions at early stages of this work. 
It is supported by the Deutsche
Forschungsgemeinschaft (DFG) through the grants GRK~2149
(Research Training Group ``Strong and Weak Interactions - from Hadrons to Dark
Matter'', S.K., F.J.\ and J.H.) and the SFB/TRR 55 (S.C.\ and W.S.).
F.J.~was supported by UKRI Future Leader Fellowship MR/T019956/1.
This project has received funding from the European Union’s Horizon Europe research and innovation programme under the Marie Sk\l{}odowska-Curie grant agreement No 101106243.
SC acknowledges support through the European Union’s Horizon 2020 research and innovation programme under the Marie Skłodowska\nobreakdash-Curie grant agreement no.~813942 (ITN EuroPLEx).
We are indebted to our colleagues in CLS for the joint
production of the $N_{\rm f}=2+1$ gauge configurations.
The authors gratefully acknowledge the \href{https://www.gauss-centre.eu}{Gauss Centre for Supercomputing (GCS)} for providing computing time through the \href{http://www.john-von-neumann-institut.de}{John von Neumann Institute for Computing (NIC)} on the supercomputer JUWELS~\cite{juwels} and, in particular, on the Booster partition of the supercomputer JURECA~\cite{jureca} at \href{http://www.fz-juelich.de/ias/jsc/}{J\"ulich Supercomputing Centre (JSC)}, and, in addition, on the supercomputer SuperMUC at the Leibniz Supercomputing Centre. GCS is the alliance of the three national supercomputing centres HLRS (Universität Stuttgart), JSC (Forschungszentrum Jülich), and LRZ (Bayerische Akademie der Wissenschaften), funded by the BMBF and the German State Ministries for Research of Baden\nobreakdash-Württemberg (MWK), Bayern (StMWFK) and Nordrhein\nobreakdash-Westfalen (MIWF). 
Additional simulations were performed on
the Regensburg Athene~2 cluster and on the SFB/TRR 55 QPACE 3 computer~\cite{Baier:2009yq,Nakamura:2011cd}.
The authors gratefully acknowledge the scientific support and HPC resources provided by the Erlangen National High Performance Computing Center (NHR@FAU) of the Friedrich-Alexander-Universität Erlangen-Nürnberg (FAU) under the NHR project \texttt{b124da}. NHR funding is provided by federal and Bavarian state authorities. NHR@FAU hardware is partially funded by the DFG – 440719683.
The two-point functions were computed using the Chroma~\cite{Edwards:2004sx}
software package, along with the locally deflated domain decomposition
solver implementation of openQCD~\cite{openQCD}, the LibHadronAnalysis
library and the multigrid solver implementation of
ref.~\cite{Heybrock:2015kpy}; additional calculations were carried out
using the code based on~\cite{mesons}.
The \texttt{pyerrors} package \cite{Joswig:2022qfe} relies on \texttt{numpy} \cite{harris2020array}
and \texttt{autograd} \cite{maclaurin2015autograd}.
Plots have been generated with \texttt{matplotlib} \cite{Hunter:2007}.

\appendix
\newpage
\section{Tables \label{a:tables}}

\begin{longtable}{clcccccccccccc}
\caption{ \label{tab:overview}
Overview of analysed ensembles along the three quark mass plane trajectories where the light equals the strange quark mass (\sym), the average bare quark mass is constant ($\Tr M$), and the strange quark mass is approximately constant ($m_{\rm s}$); ensemble name (id), (p)eriodic or (o)pen boundary conditions (bc) in time, pion and the kaon masses in physical units (see also~\cref{t:ps} for raw masses including errors) as well as the spatial~($L$) and temporal~($T$) lattice extent in physical units and the spatial extent in units of the pion mass. The lattice spacing $a$ is taken from ref.~\cite{RQCD:2022xux}. Optimized number $n_i$ of smearing iterations for light ($i=\mathrm{l}$), strange ($i=\mathrm{s}$), and charm ($i=\mathrm{c} \equiv \mathrm{c}^1=\mathrm{c}^2$) quark. Number of sources $N_{\textmd{src}}$ used to calculate point-to-all propagators on a configuration and total number of configurations $N_{\textmd{conf}}$ analysed. Analysed configurations are separated by $\Delta$ molecular dynamic units. Note that for ensembles with open boundary conditions a certain number of time slices have not been considered in the computation of some observables for sources close to the boundary due to boundary effects, see discussion in the text. Ensembles marked with a star do not enter the final extrapolation. }
  \\\toprule
traj. & id
& bc
& $\frac{m_{\pi}}{\textmd{MeV}}$
& $\frac{m_{\rm K}}{\textmd{MeV}}$
& $\frac{L}{\textmd{fm}}$
& $\frac{T}{\textmd{fm}}$
& $m_\pi L$
& $n_{\mathrm{l}}$
& $n_{\mathrm{s}}$
& $n_{\mathrm{c}}$
& $N_{\text{src}}$
& $N_{\text{conf}}$
& $\Delta$
\\
\hline\endfirsthead
\caption{ \small Overview of analysed ensembles (continued). }\\\toprule
traj. & id
& bc
& $\frac{m_{\pi}}{\textmd{MeV}}$
& $\frac{m_{\rm K}}{\textmd{MeV}}$
& $\frac{L}{\textmd{fm}}$
& $\frac{T}{\textmd{fm}}$
& $m_\pi L$
& $n_{\mathrm{l}}$
& $n_{\mathrm{s}}$
& $n_{\mathrm{c}}$
& $N_{\text{src}}$
& $N_{\text{conf}}$
& $\Delta$
\\
\hline\endhead
\hline
\multicolumn{14}{r}{\textit{Continued on next page}} \\
\endfoot
\hline
\endlastfoot
\multicolumn{14}{c}{$\beta=3.34, a = 0.098\,\mathrm{fm}$} \\\hline
\multirow{1}{*}{\sym}
           &A650& p& 370& 370& 2.34& 4.69& 4.39& 160& 160& 19& 12& 5062& 4           \\
\hline
\multirow{2}{*}{$\Tr M$}
           &A653& p& 430& 430& 2.34& 4.69& 5.10& 150& 150& 19& 12& 2525& 8           \\
           &A654& p& 335& 458& 2.34& 4.69& 3.98& 185& 165& 19& 12& 2533& 8           \\
\hline
\multicolumn{14}{c}{$\beta=3.4, a = 0.085\,\mathrm{fm}$} \\\hline
\multirow{10}{*}{$\Tr M$}
           &U103$^{\ast}$& o& 420& 420& 2.05& 10.9& 4.35& 220& 220& 25& 9& 2475& 8           \\
           &H101& o& 423& 423& 2.73& 8.19& 5.85& 220& 220& 25& 20& 2016& 4           \\
           &U102$^{\ast}$& o& 358& 445& 2.05& 10.9& 3.71& 250& 210& 25& 37& 1561& 8           \\
           &H102& o& 358& 444& 2.73& 8.19& 4.95& 250& 210& 25& 20& 1920& 4           \\
           &U101$^{\ast}$& o& 273& 465& 2.05& 10.9& 2.83& 300& 200& 25& 37& 1964& 4           \\
           &H105& o& 284& 467& 2.73& 8.19& 3.93& 300& 200& 25& 20& 1827& 4           \\
           &N101& o& 281& 467& 4.09& 10.9& 5.83& 300& 200& 25& 37& 1593& 4           \\
           &S100$^{\ast}$& o& 211& 475& 2.73& 10.9& 2.92& 350& 170& 25& 37& 983& 4           \\
           &C101& o& 220& 473& 4.09& 8.19& 4.57& 350& 170& 25& 20& 2530& 4           \\
           &D150& p& 129& 482& 5.46& 10.9& 3.56& 440& 140& 25& 32& 602& 4           \\
\hline
\multirow{3}{*}{$m_{\rm s}$}
           &H107& o& 369& 550& 2.73& 8.19& 5.10& 250& 160& 25& 18& 1064& 4           \\
           &H106& o& 276& 519& 2.73& 8.19& 3.82& 250& 160& 25& 18& 1506& 4           \\
           &C102& o& 218& 503& 4.09& 8.19& 4.52& 350& 160& 25& 18& 1496& 4           \\
\hline
\multicolumn{14}{c}{$\beta=3.46, a = 0.075\,\mathrm{fm}$} \\\hline
\multirow{1}{*}{\sym}
           &X450& p& 264& 264& 3.62& 4.83& 4.85& 400& 400& 30& 32& 1059& 4           \\
\hline
\multirow{5}{*}{$\Tr M$}
           &B450& p& 421& 421& 2.42& 4.83& 5.15& 270& 270& 30& 16& 1612& 4           \\
           &S400& o& 354& 445& 2.42& 9.66& 4.33& 310& 260& 30& 21& 2872& 4           \\
           &N451& p& 289& 466& 3.62& 9.66& 5.31& 375& 250& 30& 32& 1011& 4           \\
           &D450& p& 216& 480& 4.83& 9.66& 5.29& 480& 200& 30& 32& 544& 4           \\
           &D452& p& 155& 487& 4.83& 9.66& 3.80& 530& 180& 30& 32& 995& 4           \\
\hline
\multirow{3}{*}{$m_{\rm s}$}
           &B452$^{\ast}$& p& 352& 548& 2.42& 4.83& 4.31& 310& 200& 30& 16& 1944& 4           \\*
           &N450& p& 287& 527& 3.62& 9.66& 5.27& 375& 200& 30& 32& 1130& 4           \\*
           &D451& p& 218& 507& 4.83& 9.66& 5.34& 480& 200& 30& 32& 1028& 4           \\
\hline
\multicolumn{14}{c}{$\beta=3.55, a = 0.064\,\mathrm{fm}$} \\\hline
\multirow{3}{*}{\sym}
           &X250& p& 348& 348& 3.06& 4.08& 5.41& 445& 445& 45& 32& 1500& 4           \\
           &X251& p& 267& 267& 3.06& 4.08& 4.15& 540& 540& 45& 32& 1411& 4           \\
           &D250& p& 199& 199& 4.08& 8.17& 4.12& 660& 660& 45& 32& 458& 4           \\
\hline
\multirow{9}{*}{$\Tr M$}
           &N202& o& 414& 414& 3.06& 8.17& 6.42& 390& 390& 45& 20& 899& 4           \\
           &N203& o& 348& 445& 3.06& 8.17& 5.40& 445& 375& 45& 21& 1543& 4           \\
           &S201$^{\ast}$& o& 294& 471& 2.04& 8.17& 3.04& 540& 360& 45& 20& 1955& 8           \\
           &X252$^{\ast}$& p& 288& 467& 2.30& 8.17& 3.35& 540& 360& 45& 32& 3508& 4           \\
           &X253$^{\ast}$& p& 287& 466& 2.55& 8.17& 3.71& 540& 360& 45& 32& 3005& 4           \\
           &N200& o& 286& 466& 3.06& 8.17& 4.45& 540& 360& 45& 21& 1712& 4           \\
           &D251& p& 286& 465& 4.08& 8.17& 5.91& 540& 360& 45& 32& 403& 4           \\
           &D200& o& 201& 481& 4.08& 8.17& 4.16& 660& 290& 45& 20& 1999& 4           \\
           &E250& p& 130& 493& 6.13& 12.3& 4.05& 795& 285& 45& 32& 503& 4           \\
\hline
\multirow{3}{*}{$m_{\rm s}$}
           &N204& o& 353& 548& 3.06& 8.17& 5.48& 445& 285& 45& 24& 1500& 4           \\
           &N201& o& 287& 527& 3.06& 8.17& 4.45& 540& 285& 45& 24& 1522& 4           \\
           &D201& o& 204& 506& 4.08& 8.17& 4.21& 660& 285& 45& 20& 1078& 4           \\
\hline
\multicolumn{14}{c}{$\beta=3.7, a = 0.049\,\mathrm{fm}$} \\\hline
\multirow{1}{*}{\sym}
           &N306& o& 343& 343& 2.37& 6.32& 4.12& 750& 750& 75& 21& 1507& 4           \\
\hline
\multirow{4}{*}{$\Tr M$}
           &N300$^{\ast}$& o& 426& 426& 2.37& 6.32& 5.11& 640& 640& 75& 18& 1540& 4           \\
           &N302& o& 350& 454& 2.37& 6.32& 4.20& 750& 620& 75& 21& 2201& 4           \\
           &J303& o& 259& 478& 3.16& 9.48& 4.14& 950& 525& 75& 20& 999& 8           \\
           &E300& o& 175& 495& 4.74& 9.48& 4.20& 800& 310& 55& 16& 1137& 4           \\
\hline
\multirow{2}{*}{$m_{\rm s}$}
           &N304& o& 355& 556& 2.37& 6.32& 4.26& 750& 465& 75& 24& 1651& 4           \\
           &J304& o& 261& 527& 3.16& 9.48& 4.18& 950& 465& 75& 3& 1521& 4           \\
\hline
\multicolumn{14}{c}{$\beta=3.85, a = 0.039\,\mathrm{fm}$} \\\hline
\multirow{2}{*}{$\Tr M$}
           &J500& o& 414& 414& 2.48& 7.44& 5.20& 1000& 1000& 115& 3& 1837& 8           \\
           &J501& o& 337& 448& 2.48& 7.44& 4.23& 1225& 1025& 115& 20& 2292& 4           \\
\bottomrule
\end{longtable}

\newpage

\begin{longtable}{cllllll}
\caption{ \label{t:hopping}Hopping parameters used in the computation of the correlation functions. $\kappa_{\rm l}$ and $\kappa_{\rm s}$ match those that have been used in the generation of the ensemble. Ensembles marked with a star do not enter the final extrapolation. }
  \\\toprule
traj. & id
& \multicolumn{1}{c}{$\kappa_{\rm l}$}
& \multicolumn{1}{c}{$\kappa_{\rm s}$}
& \multicolumn{1}{c}{$\kappa_{\rm c^1}$}
& \multicolumn{1}{c}{$\kappa_{\rm c^2}$}
\\
\hline\endfirsthead
\caption{ Overview of hopping parameters (continued). }\\\toprule
traj. & id
& \multicolumn{1}{c}{$\kappa_{\rm l}$}
& \multicolumn{1}{c}{$\kappa_{\rm s}$}
& \multicolumn{1}{c}{$\kappa_{\rm c^1}$}
& \multicolumn{1}{c}{$\kappa_{\rm c^2}$}
\\
\hline\endhead
\hline
\multicolumn{6}{r}{\textit{Continued on next page}} \\
\endfoot
\hline
\endlastfoot
\multicolumn{6}{c}{$\beta=3.34, a = 0.098\,\mathrm{fm}$} \\\hline
\multirow{1}{*}{\sym}
           &A650& 0.1366& 0.1366& 0.121904& 0.120692           \\
\hline
\multirow{2}{*}{$\Tr M$}
           &A653& 0.1365715& 0.1365715& 0.121904& 0.120692           \\
           &A654& 0.13675& 0.136216193& 0.121904& 0.120692           \\
\hline
\multicolumn{6}{c}{$\beta=3.4, a = 0.085\,\mathrm{fm}$} \\\hline
\multirow{10}{*}{$\Tr M$}
           &U103$^{\ast}$& 0.13675962& 0.13675962& 0.124056& 0.123147           \\
           &H101& 0.13675962& 0.13675962& 0.124056& 0.123147           \\
           &U102$^{\ast}$& 0.136865& 0.136549339& 0.124056& 0.123147           \\
           &H102& 0.136865& 0.136549339& 0.124056& 0.123147           \\
           &U101$^{\ast}$& 0.13697& 0.13634079& 0.124056& 0.123147           \\
           &H105& 0.13697& 0.13634079& 0.124056& 0.123147           \\
           &N101& 0.13697& 0.13634079& 0.124056& 0.123147           \\
           &S100$^{\ast}$& 0.13703& 0.136222041& 0.124056& 0.123147           \\
           &C101& 0.13703& 0.136222041& 0.124056& 0.123147           \\
           &D150& 0.137088& 0.13610755& 0.124056& 0.123147           \\
\hline
\multirow{3}{*}{$m_{\rm s}$}
           &H107& 0.13694566590798& 0.136203165143476& 0.124056& 0.123147           \\
           &H106& 0.137015570024& 0.136148704478& 0.124056& 0.123147           \\
           &C102& 0.13705084580022& 0.13612906255557& 0.124056& 0.123147           \\
\hline
\multicolumn{6}{c}{$\beta=3.46, a = 0.075\,\mathrm{fm}$} \\\hline
\multirow{1}{*}{\sym}
           &X450& 0.136994& 0.136994& 0.1254266& 0.12617097           \\
\hline
\multirow{5}{*}{$\Tr M$}
           &B450& 0.13689& 0.13689& 0.1258025& 0.12503           \\
           &S400& 0.136984& 0.136702387& 0.126983423& 0.125563292           \\
           &N451& 0.1370616& 0.1365480771& 0.126983423& 0.125563292           \\
           &D450& 0.137126& 0.136420428639937& 0.126983423& 0.125563292           \\
           &D452& 0.137163675& 0.136345904546& 0.126983423& 0.125563292           \\
\hline
\multirow{3}{*}{$m_{\rm s}$}
           &B452$^{\ast}$& 0.1370455& 0.136378044& 0.1258025& 0.12503           \\
           &N450& 0.1370986& 0.136352601& 0.1258025& 0.12503           \\
           &D451& 0.13714& 0.136337761& 0.126983423& 0.125563292           \\
\hline
\multicolumn{6}{c}{$\beta=3.55, a = 0.064\,\mathrm{fm}$} \\\hline
\multirow{3}{*}{\sym}
           &X250& 0.13705& 0.13705& 0.12751596& 0.12871743           \\
           &X251& 0.1371& 0.1371& 0.12765893& 0.12884069           \\
           &D250& 0.13713129& 0.13713129& 0.12884069& 0.12765893           \\
\hline
\multirow{9}{*}{$\Tr M$}
           &N202& 0.137& 0.137& 0.128651119& 0.1274374           \\*
           &N203& 0.13708& 0.136840284& 0.128651119& 0.1274374           \\*
           &S201$^{\ast}$& 0.13714& 0.13672086& 0.128651119& 0.1274374           \\*
           &X252$^{\ast}$& 0.13714& 0.13672086& 0.128651119& 0.1274374           \\*
           &X253$^{\ast}$& 0.13714& 0.13672086& 0.128651119& 0.1274374           \\*
           &N200& 0.13714& 0.13672086& 0.128651119& 0.1274374           \\*
           &D251& 0.13714& 0.13672086& 0.128651119& 0.1274374           \\*
           &D200& 0.1372& 0.136601748& 0.128651119& 0.1274374           \\*
           &E250& 0.137232867& 0.136536633& 0.128651119& 0.1274374           \\
\hline
\multirow{3}{*}{$m_{\rm s}$}
           &N204& 0.137112& 0.136575049& 0.128651119& 0.1274374           \\
           &N201& 0.13715968& 0.136561319& 0.128651119& 0.1274374           \\
           &D201& 0.1372067& 0.136546844& 0.128651119& 0.1274374           \\
\hline
\multicolumn{6}{c}{$\beta=3.7, a = 0.049\,\mathrm{fm}$} \\\hline
\multirow{1}{*}{\sym}
           &N306& 0.13705013& 0.13705013& 0.13062697& 0.13018588           \\
\hline
\multirow{4}{*}{$\Tr M$}
           &N300$^{\ast}$& 0.137& 0.137& 0.13062697& 0.13018588           \\
           &N302& 0.137064& 0.1368721791358& 0.13062697& 0.13018588           \\
           &J303& 0.137123& 0.1367546608& 0.13062697& 0.13018588           \\
           &E300& 0.137163& 0.136675163617733& 0.13062697& 0.13018588           \\
\hline
\multirow{2}{*}{$m_{\rm s}$}
           &N304& 0.137079325093654& 0.136665430105663& 0.13062697& 0.13018588           \\
           &J304& 0.13713& 0.1366569203& 0.13062697& 0.13018588           \\
\hline
\multicolumn{6}{c}{$\beta=3.85, a = 0.039\,\mathrm{fm}$} \\\hline
\multirow{2}{*}{$\Tr M$}
           &J500& 0.136852& 0.136852& 0.13242984& 0.13206693           \\
           &J501& 0.1369032& 0.136749715& 0.13242984& 0.13206693           \\
\bottomrule
\end{longtable}


\begin{longtable}{cllllll}
\caption{ \label{t:ps}The gluonic observable $t_0/a^2$ and the pseudoscalar masses that are used in the extrapolation to the physical point in lattice units. Ensembles marked with a star do not enter the final extrapolation. }
  \\\toprule
traj. & id
& $t_0/a^2$
& $am_\pi$
& $am_{\rm K}$
& $am_{\bar{\rm D}^1}$
& $am_{\bar{\rm D}^2}$
\\
\hline\endfirsthead
\caption{ Overview of $t_0/a^2$ and pseudoscalar masses (continued). }\\\toprule
traj. & id
& $t_0/a^2$
& $am_\pi$
& $am_{\rm K}$
& $am_{\bar{\rm D}^1}$
& $am_{\bar{\rm D}^2}$
\\
\hline\endhead
\hline
\multicolumn{7}{r}{\textit{Continued on next page}} \\
\endfoot
\hline
\endlastfoot
\multicolumn{7}{c}{$\beta=3.34, a = 0.098\,\mathrm{fm}$} \\\hline
\multirow{1}{*}{\sym}
           &A650& 2.2860(76)& 0.1829(13)& 0.1829(13)& 0.90948(53)& 0.94916(52)           \\
\hline
\multirow{2}{*}{$\Tr M$}
           &A653& 2.1727(62)& 0.2125(10)& 0.2125(10)& 0.91999(51)& 0.95951(50)           \\
           &A654& 2.1932(83)& 0.1657(14)& 0.2268(11)& 0.91846(53)& 0.95798(55)           \\
\hline \pagebreak[4]
\multicolumn{7}{c}{$\beta=3.4, a = 0.085\,\mathrm{fm}$} \\* \hline
\multirow{10}{*}{$\Tr M$}
           &U103$^{\ast}$& 2.8841(63)& 0.18138(75)& 0.18138(75)& 0.82103(62)& 0.85207(65)           \\*
           &H101& 2.8468(56)& 0.18285(67)& 0.18285(67)& 0.82029(49)& 0.85132(51)           \\*
           &U102$^{\ast}$& 2.888(12)& 0.1546(11)& 0.19240(78)& 0.81953(60)& 0.85049(63)           \\*
           &H102& 2.8806(75)& 0.15458(66)& 0.19208(74)& 0.81819(60)& 0.84913(63)           \\*
           &U101$^{\ast}$& 2.926(13)& 0.1178(25)& 0.2011(13)& 0.81845(73)& 0.84946(78)           \\*
           &H105& 2.894(13)& 0.1227(10)& 0.20196(76)& 0.81686(75)& 0.84771(78)           \\*
           &N101& 2.8930(32)& 0.12153(53)& 0.20172(31)& 0.81758(27)& 0.84850(29)           \\*
           &S100$^{\ast}$& 2.9223(89)& 0.0914(40)& 0.20528(75)& 0.81651(71)& 0.84742(77)           \\*
           &C101& 2.9105(47)& 0.0952(19)& 0.2043(14)& 0.81666(51)& 0.84754(54)           \\*
           &D150& 2.9475(34)& 0.0557(11)& 0.20840(29)& 0.81425(54)& 0.84513(57)           \\*
\hline
\multirow{3}{*}{$m_{\rm s}$}
           &H107& 2.716(11)& 0.1595(11)& 0.23777(89)& 0.83190(71)& 0.86265(76)           \\
           &H106& 2.8208(58)& 0.1193(16)& 0.22428(78)& 0.82381(66)& 0.85469(68)           \\
           &C102& 2.8684(53)& 0.0942(20)& 0.2173(13)& 0.81931(66)& 0.85015(70)           \\
\hline
\multicolumn{7}{c}{$\beta=3.46, a = 0.075\,\mathrm{fm}$} \\\hline
\multirow{1}{*}{\sym}
           &X450& 3.9901(85)& 0.10098(43)& 0.10098(43)& 0.73124(40)& 0.70450(33)           \\
\hline
\multirow{5}{*}{$\Tr M$}
           &B450& 3.663(13)& 0.16105(51)& 0.16105(51)& 0.73582(45)& 0.76313(44)           \\
           &S400& 3.6917(77)& 0.13540(42)& 0.17020(40)& 0.69167(36)& 0.74324(37)           \\
           &N451& 3.6821(66)& 0.11069(42)& 0.17826(25)& 0.69183(24)& 0.74340(26)           \\
           &D450& 3.7068(57)& 0.08270(48)& 0.18355(19)& 0.69092(32)& 0.74249(36)           \\
           &D452& 3.7266(36)& 0.05939(61)& 0.18638(15)& 0.68970(31)& 0.74128(35)           \\
\hline
\multirow{3}{*}{$m_{\rm s}$}
           &B452$^{\ast}$& 3.5284(69)& 0.13468(48)& 0.20969(35)& 0.74516(32)& 0.77239(33)           \\
           &N450& 3.5919(47)& 0.10972(35)& 0.20177(23)& 0.74035(25)& 0.76756(28)           \\
           &D451& 3.6645(52)& 0.08337(27)& 0.19381(16)& 0.69308(25)& 0.74461(27)           \\
\hline
\multicolumn{7}{c}{$\beta=3.55, a = 0.064\,\mathrm{fm}$} \\\hline
\multirow{3}{*}{\sym}
           &X250& 5.321(16)& 0.11261(28)& 0.11261(28)& 0.63537(30)& 0.58974(31)           \\
           &X251& 5.499(10)& 0.08642(34)& 0.08642(34)& 0.62294(43)& 0.57781(43)           \\
           &D250& 5.619(14)& 0.06444(53)& 0.06444(53)& 0.57317(40)& 0.61829(45)           \\
\hline
\multirow{9}{*}{$\Tr M$}
           &N202& 5.166(21)& 0.13382(48)& 0.13382(48)& 0.59858(64)& 0.64445(63)           \\
           &N203& 5.1466(73)& 0.11259(32)& 0.14397(30)& 0.59821(31)& 0.64396(32)           \\
           &S201$^{\ast}$& 5.164(11)& 0.09515(57)& 0.15232(51)& 0.59885(71)& 0.64458(77)           \\
           &X252$^{\ast}$& 5.160(10)& 0.09317(27)& 0.15106(21)& 0.59868(25)& 0.64448(27)           \\
           &X253$^{\ast}$& 5.1535(69)& 0.09276(21)& 0.15075(19)& 0.59861(20)& 0.64449(22)           \\
           &N200& 5.1633(70)& 0.09264(46)& 0.15069(38)& 0.59830(33)& 0.64407(36)           \\
           &D251& 5.1527(83)& 0.09235(27)& 0.15053(27)& 0.59815(22)& 0.64393(24)           \\
           &D200& 5.1790(62)& 0.06503(57)& 0.15558(56)& 0.59766(28)& 0.64344(31)           \\
           &E250& 5.2027(44)& 0.04218(25)& 0.159369(67)& 0.59714(35)& 0.64299(38)           \\
\hline
\multirow{3}{*}{$m_{\rm s}$}
           &N204& 4.947(10)& 0.11413(47)& 0.17729(42)& 0.60767(43)& 0.65342(46)           \\*
           &N201& 5.0426(78)& 0.09277(44)& 0.17042(30)& 0.60324(30)& 0.64898(33)           \\*
           &D201& 5.1363(84)& 0.06582(58)& 0.16364(37)& 0.59901(37)& 0.64478(39)           \\*
\hline
\multicolumn{7}{c}{$\beta=3.7, a = 0.049\,\mathrm{fm}$} \\\hline
\multirow{1}{*}{\sym}
           &N306& 8.811(48)& 0.08575(72)& 0.08575(72)& 0.47077(60)& 0.48866(61)           \\
\hline
\multirow{4}{*}{$\Tr M$}
           &N300$^{\ast}$& 8.566(39)& 0.10649(45)& 0.10649(45)& 0.47366(65)& 0.49153(68)           \\
           &N302& 8.526(25)& 0.08758(52)& 0.11370(49)& 0.47624(45)& 0.49409(46)           \\
           &J303& 8.620(15)& 0.06476(26)& 0.11963(20)& 0.47466(28)& 0.49250(29)           \\
           &E300& 8.6193(63)& 0.04371(35)& 0.12395(32)& 0.47422(24)& 0.49204(25)           \\
\hline
\multirow{2}{*}{$m_{\rm s}$}
           &N304& 8.328(25)& 0.08873(76)& 0.13923(68)& 0.48117(62)& 0.49899(63)           \\
           &J304& 8.501(15)& 0.06537(27)& 0.13175(28)& 0.47669(34)& 0.49447(36)           \\
\hline
\multicolumn{7}{c}{$\beta=3.85, a = 0.039\,\mathrm{fm}$} \\\hline
\multirow{2}{*}{$\Tr M$}
           &J500& 13.973(28)& 0.08127(30)& 0.08127(30)& 0.35574(43)& 0.37159(46)           \\
           &J501& 14.007(70)& 0.06617(35)& 0.08801(33)& 0.35616(31)& 0.37206(32)           \\
\bottomrule
\end{longtable}

\begin{longtable}{clllll}
\caption{ \label{t:decayc}Bare decay constants in lattice units based on $\mathrm{O}(a)$ improved currents defined in~\cref{eq:impc}. Ensembles marked with a star do not enter the final extrapolation. }
  \\\toprule
traj. & id
& $af_{\rm D^1}$
& $af_{\rm D^2}$
& $af_{\rm D_{\rm s}^1}$
& $af_{\rm D_{\rm s}^2}$
\\
\hline\endfirsthead
\caption{ Overview of bare decay constants (continued). }\\\toprule
traj. & id
& $af_{\rm D^1}$
& $af_{\rm D^2}$
& $af_{\rm D_{\rm s}^1}$
& $af_{\rm D_{\rm s}^2}$
\\
\hline\endhead
\hline
\multicolumn{6}{r}{\textit{Continued on next page}} \\
\endfoot
\hline
\endlastfoot
\multicolumn{6}{c}{$\beta=3.34, a = 0.098\,\mathrm{fm}$} \\\hline
\multirow{1}{*}{\sym}
           &A650& 0.10683(24)& 0.10540(24)& 0.10683(24)& 0.10540(24)           \\
\hline
\multirow{2}{*}{$\Tr M$}
           &A653& 0.11141(25)& 0.10990(26)& 0.11141(25)& 0.10990(26)           \\
           &A654& 0.10797(38)& 0.10644(40)& 0.11477(25)& 0.11326(27)           \\
\hline
\multicolumn{6}{c}{$\beta=3.4, a = 0.085\,\mathrm{fm}$} \\\hline
\multirow{10}{*}{$\Tr M$}
           &U103$^{\ast}$& 0.10203(47)& 0.10121(55)& 0.10203(47)& 0.10121(55)           \\
           &H101& 0.10186(33)& 0.10086(35)& 0.10186(33)& 0.10086(35)           \\
           &U102$^{\ast}$& 0.09913(47)& 0.09814(49)& 0.10362(31)& 0.10261(33)           \\
           &H102& 0.09863(55)& 0.09761(58)& 0.10331(36)& 0.10229(38)           \\
           &U101$^{\ast}$& 0.09599(69)& 0.09507(72)& 0.10540(34)& 0.10440(35)           \\
           &H105& 0.09581(64)& 0.09470(67)& 0.10570(28)& 0.10468(29)           \\
           &N101& 0.09683(33)& 0.09582(39)& 0.10603(18)& 0.10502(18)           \\
           &S100$^{\ast}$& 0.09424(75)& 0.09325(78)& 0.10707(23)& 0.10606(24)           \\
           &C101& 0.09501(67)& 0.09393(71)& 0.10716(16)& 0.10613(16)           \\
           &D150& 0.09148(64)& 0.09042(67)& 0.10732(22)& 0.10628(23)           \\
\hline
\multirow{3}{*}{$m_{\rm s}$}
           &H107& 0.10203(75)& 0.10099(78)& 0.11166(44)& 0.11064(45)           \\
           &H106& 0.09802(54)& 0.09701(56)& 0.10933(37)& 0.10828(39)           \\
           &C102& 0.09531(74)& 0.09422(79)& 0.10916(25)& 0.10814(26)           \\
\hline \pagebreak[4]
\multicolumn{6}{c}{$\beta=3.46, a = 0.075\,\mathrm{fm}$} \\\hline
\multirow{1}{*}{\sym}
           &X450& 0.08331(31)& 0.08402(28)& 0.08331(31)& 0.08402(28)           \\
\hline
\multirow{5}{*}{$\Tr M$}
           &B450& 0.09228(26)& 0.09157(25)& 0.09228(26)& 0.09157(25)           \\
           &S400& 0.09138(29)& 0.09007(31)& 0.09594(19)& 0.09465(20)           \\
           &N451& 0.09029(24)& 0.08898(29)& 0.098113(93)& 0.09682(11)           \\
           &D450& 0.08827(38)& 0.08692(43)& 0.09971(13)& 0.09844(14)           \\
           &D452& 0.08643(38)& 0.08518(46)& 0.10001(13)& 0.09872(14)           \\
\hline
\multirow{3}{*}{$m_{\rm s}$}
           &B452$^{\ast}$& 0.09352(31)& 0.09281(33)& 0.10232(14)& 0.10156(14)           \\
           &N450& 0.08994(28)& 0.08918(29)& 0.10115(13)& 0.10037(12)           \\
           &D451& 0.08855(29)& 0.08717(32)& 0.10098(11)& 0.09969(11)           \\
\hline
\multicolumn{6}{c}{$\beta=3.55, a = 0.064\,\mathrm{fm}$} \\\hline
\multirow{3}{*}{\sym}
           &X250& 0.07693(18)& 0.07785(17)& 0.07693(18)& 0.07785(17)           \\
           &X251& 0.07285(28)& 0.07380(27)& 0.07285(28)& 0.07380(27)           \\
           &D250& 0.07101(24)& 0.07006(26)& 0.07101(24)& 0.07006(26)           \\
\hline
\multirow{9}{*}{$\Tr M$}
           &N202& 0.08114(31)& 0.08024(33)& 0.08114(31)& 0.08024(33)           \\
           &N203& 0.07947(25)& 0.07846(27)& 0.08356(16)& 0.08259(18)           \\
           &S201$^{\ast}$& 0.07655(86)& 0.07551(91)& 0.08515(33)& 0.08419(38)           \\
           &X252$^{\ast}$& 0.07775(23)& 0.07679(25)& 0.08523(12)& 0.08430(13)           \\
           &X253$^{\ast}$& 0.07806(19)& 0.07718(23)& 0.08526(11)& 0.08437(10)           \\
           &N200& 0.07788(29)& 0.07686(34)& 0.08542(15)& 0.08449(16)           \\
           &D251& 0.07795(22)& 0.07694(24)& 0.08521(17)& 0.08427(18)           \\
           &D200& 0.07604(33)& 0.07501(40)& 0.08690(15)& 0.08599(17)           \\
           &E250& 0.07434(40)& 0.07338(46)& 0.08756(14)& 0.08668(16)           \\
\hline
\multirow{3}{*}{$m_{\rm s}$}
           &N204& 0.08241(55)& 0.08148(60)& 0.09019(46)& 0.08937(46)           \\
           &N201& 0.07901(34)& 0.07802(40)& 0.08910(15)& 0.08824(16)           \\
           &D201& 0.07651(43)& 0.07553(51)& 0.08799(18)& 0.08706(20)           \\
\hline
\multicolumn{6}{c}{$\beta=3.7, a = 0.049\,\mathrm{fm}$} \\\hline
\multirow{1}{*}{\sym}
           &N306& 0.06244(47)& 0.06226(49)& 0.06244(47)& 0.06226(49)           \\
\hline
\multirow{4}{*}{$\Tr M$}
           &N300$^{\ast}$& 0.06374(43)& 0.06345(45)& 0.06374(43)& 0.06345(45)           \\
           &N302& 0.06398(37)& 0.06377(39)& 0.06735(25)& 0.06714(26)           \\
           &J303& 0.06169(35)& 0.06148(37)& 0.06862(18)& 0.06842(19)           \\
           &E300& 0.06019(27)& 0.05994(24)& 0.06956(17)& 0.06934(17)           \\
\hline
\multirow{2}{*}{$m_{\rm s}$}
           &N304& 0.06505(53)& 0.06485(54)& 0.07194(40)& 0.07176(39)           \\
           &J304& 0.06138(36)& 0.06111(37)& 0.07078(23)& 0.07061(24)           \\
\hline
\multicolumn{6}{c}{$\beta=3.85, a = 0.039\,\mathrm{fm}$} \\\hline
\multirow{2}{*}{$\Tr M$}
           &J500& 0.05136(42)& 0.05128(44)& 0.05136(42)& 0.05128(44)           \\
           &J501& 0.05045(26)& 0.05036(25)& 0.05320(20)& 0.05315(21)           \\
\bottomrule
\end{longtable}

\begin{table}[htb]
\begin{center}
  {\scriptsize
    \begin{tabular}{ccccccc}
      \toprule
$\beta$&3.34 & 3.4 &            3.46&            3.55&          3.7&           3.85\\\midrule
$Z_{\rm A}$~\cite{DallaBrida:2018tpn} & 0.7510(11) & 0.75629(65) & 0.76172(39) & 0.76994(34) & 0.78356(32) & 0.79675(45) \\ 
$b_{\rm A}$~\cite{Bali:2021qem} & 1.249(16) & 1.244(16) & 1.239(15) & 1.232(15) & 1.221(13) & 1.211(12) \\
$c_{\rm A}$~\cite{Bulava:2015bxa} & $-0.055709847$ & $-0.048973873$ & $-0.04320929$ & $-0.036074160$ & $-0.027286251$ & $-0.021260222$ \\ 
$\kappa_{\rm crit}$~\cite{RQCD:2022xux} & 0.1366938(45) & 0.1369153(9) & 0.1370613(10) & 0.1371715(10) & 0.1371530(9) & 0.1369767(26)\\ \bottomrule
  \end{tabular}}
\end{center}
\caption{\label{tab:inputparam}
Summary table of input parameters used for the calculation of the decay constants. Note that $c_{\rm A}$ defines the valence action, so no uncertainty enters for this quantity. For $\kappa_{\rm crit}$ we use the values in ref.~\cite{RQCD:2022xux} labeled $\kappa_{\rm crit}$ (int), and $b_{\rm A}$ from eq.~(5.4) in ref.~\cite{Bali:2021qem} is used. Note that we set $\bar{b}_{\rm A}$ to zero because all coefficients in eq.~(5.3) in ref.~\cite{Bali:2021qem} are compatible with zero.
}
\end{table}

\newpage
\section{Additional figures \label{a:plots}}

\begin{figure}[ht]
	\centering
	\includegraphics[width=0.5\textwidth]{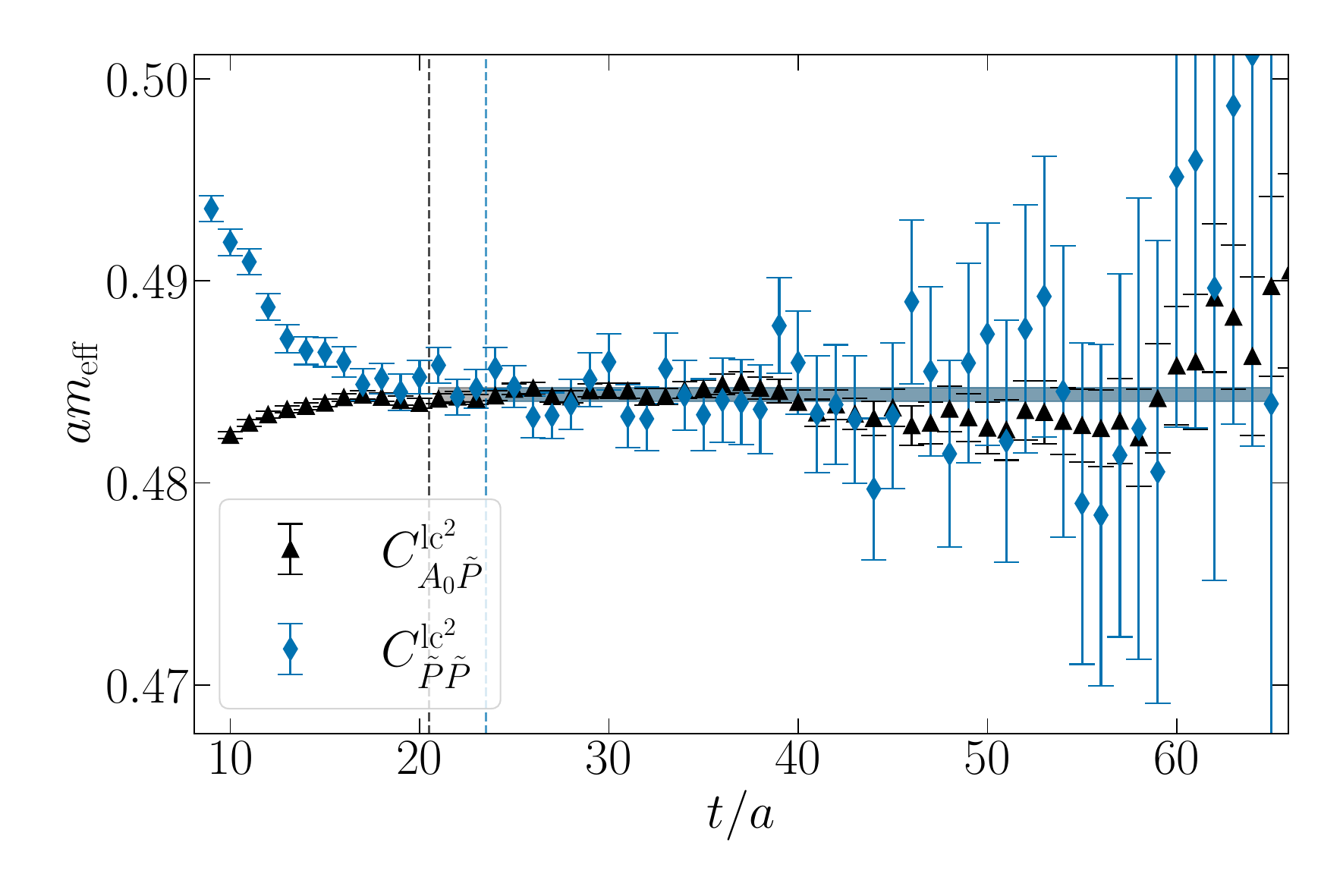}%
	\includegraphics[width=0.5\textwidth]{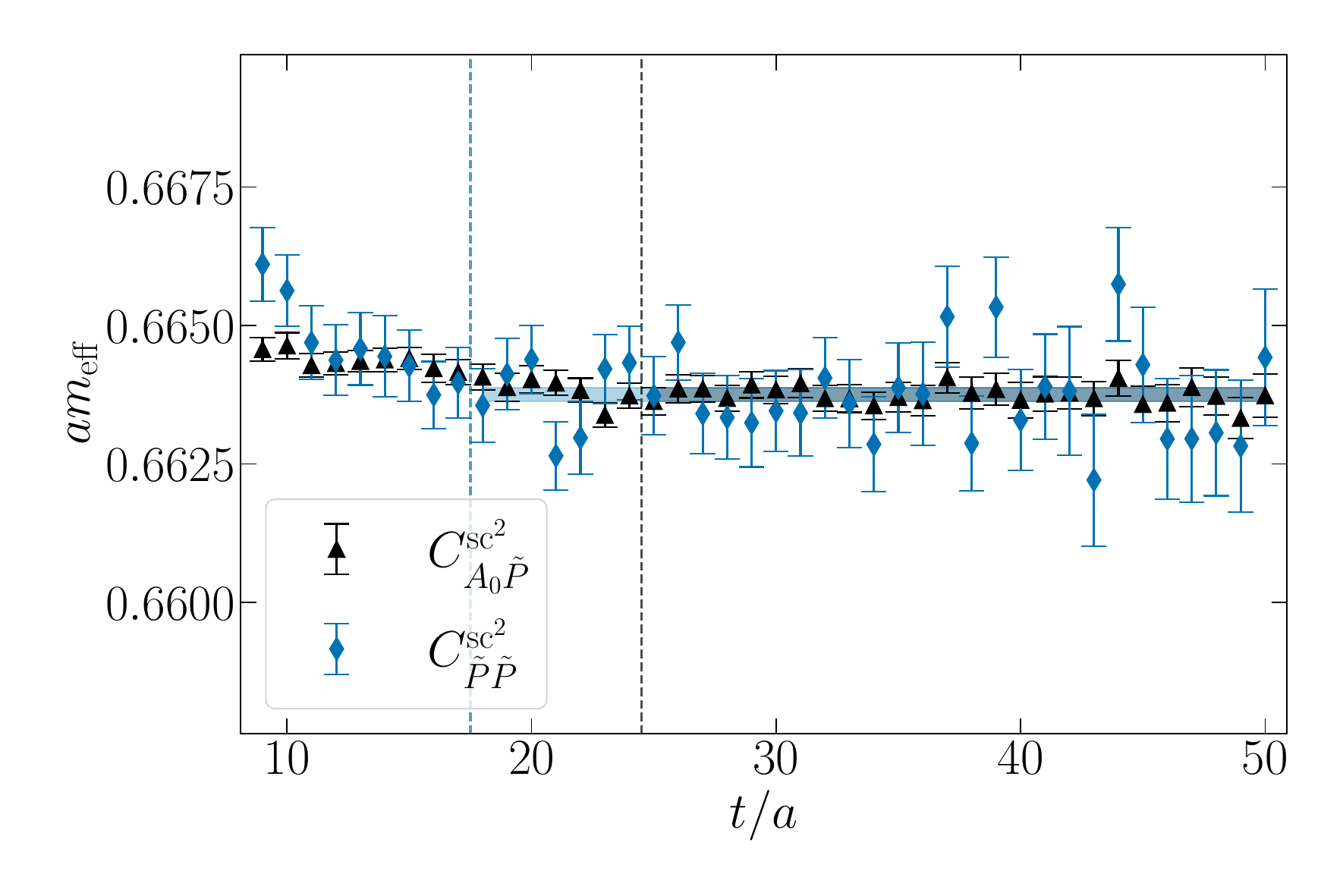}\\
	\includegraphics[width=0.5\textwidth]{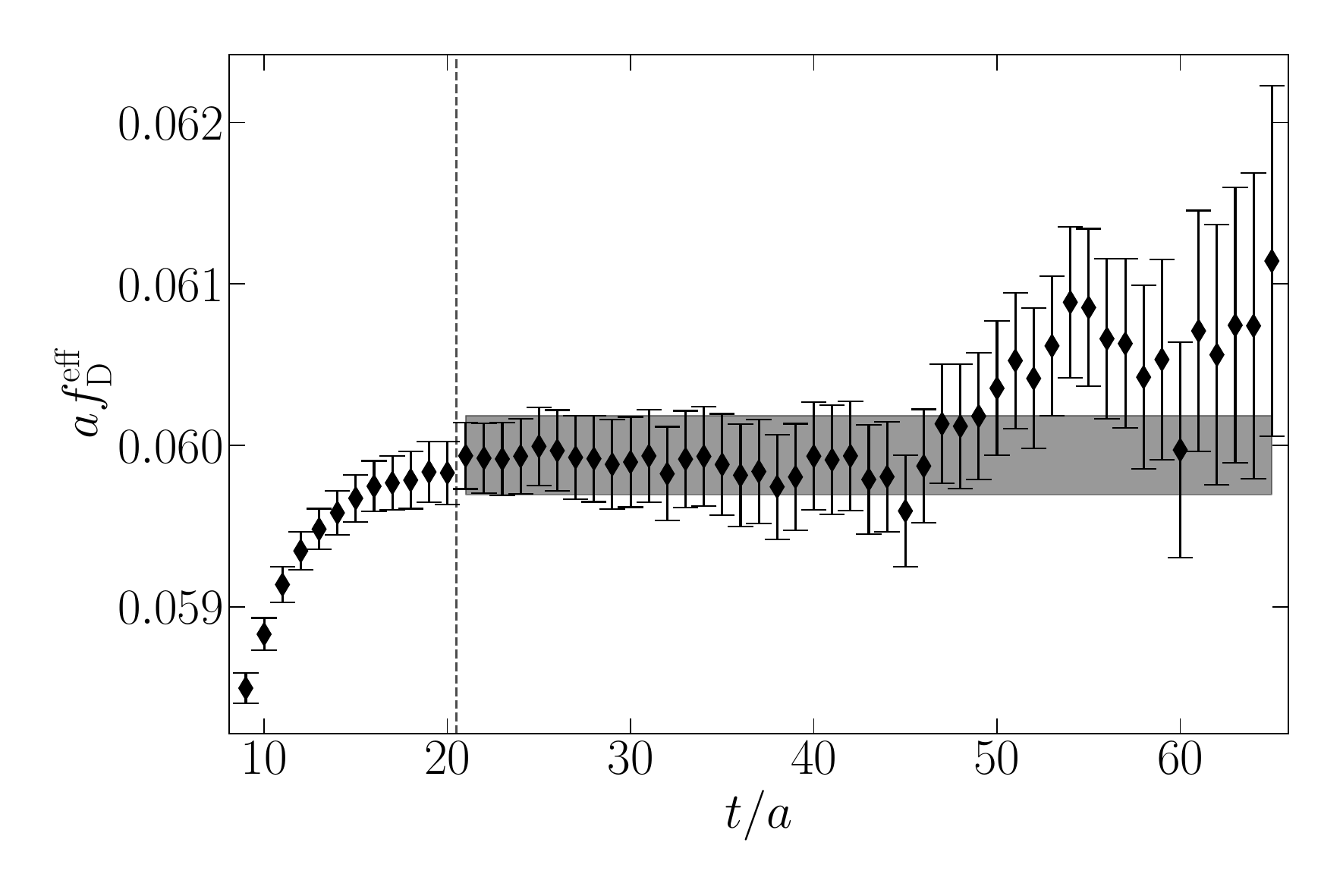}%
	\includegraphics[width=0.5\textwidth]{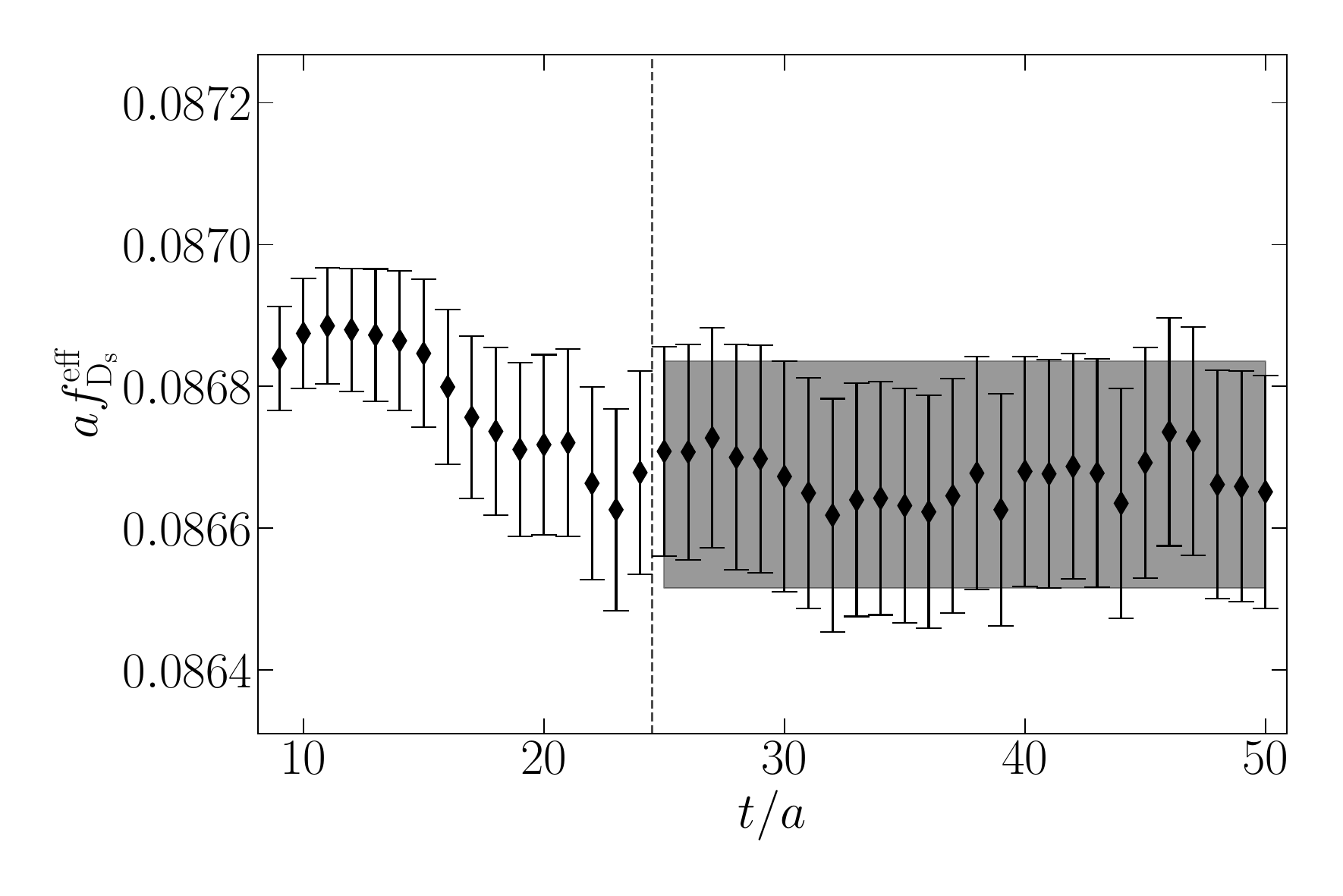}%
	\caption{Results from our combined fits to the heavy-light current on E300 (left, $m_\pi=175$\,MeV, $a=0.049\,$fm) and the heavy-strange current on E250 (right, $m_\pi=130$\,MeV, $a=0.064\,$fm) as in figure \ref{f:simfit}. 
	}
	\label{f:simfit_extra}
\end{figure}

\begin{figure}[ht]
	\centerline{
	\includegraphics[width=0.5\textwidth]{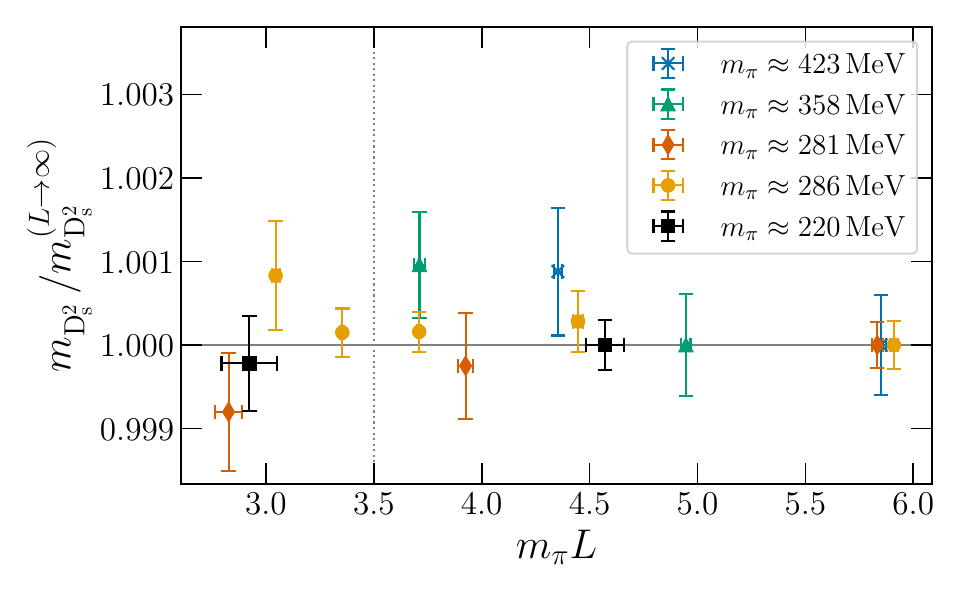}
	\includegraphics[width=0.5\textwidth]{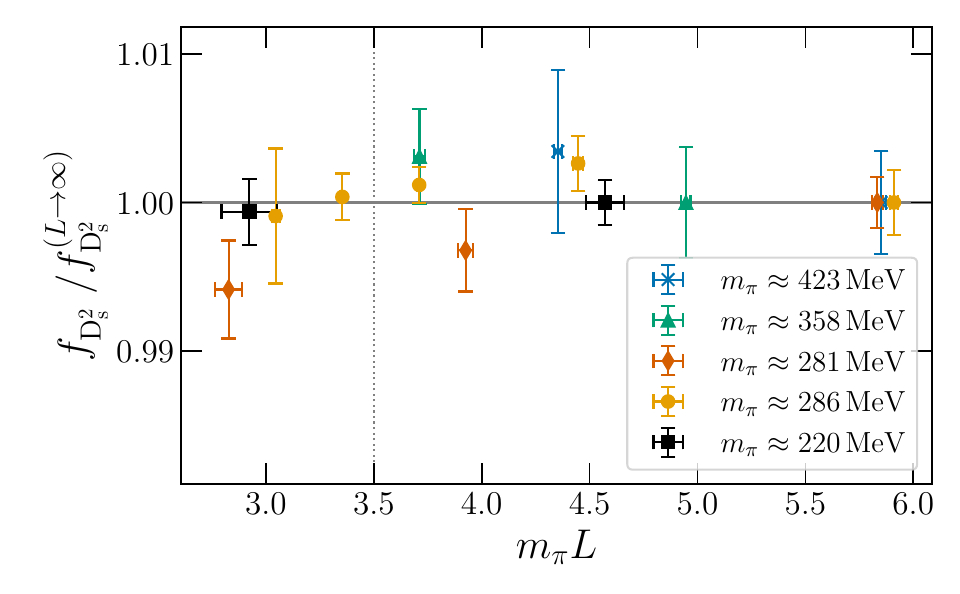}}
	\centerline{
	\includegraphics[width=0.5\textwidth]{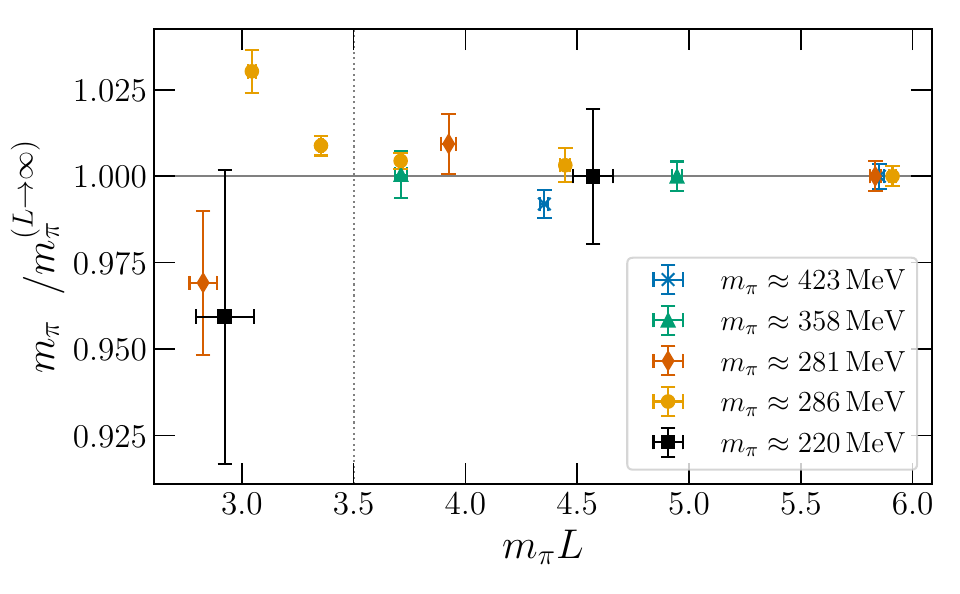}
	\includegraphics[width=0.5\textwidth]{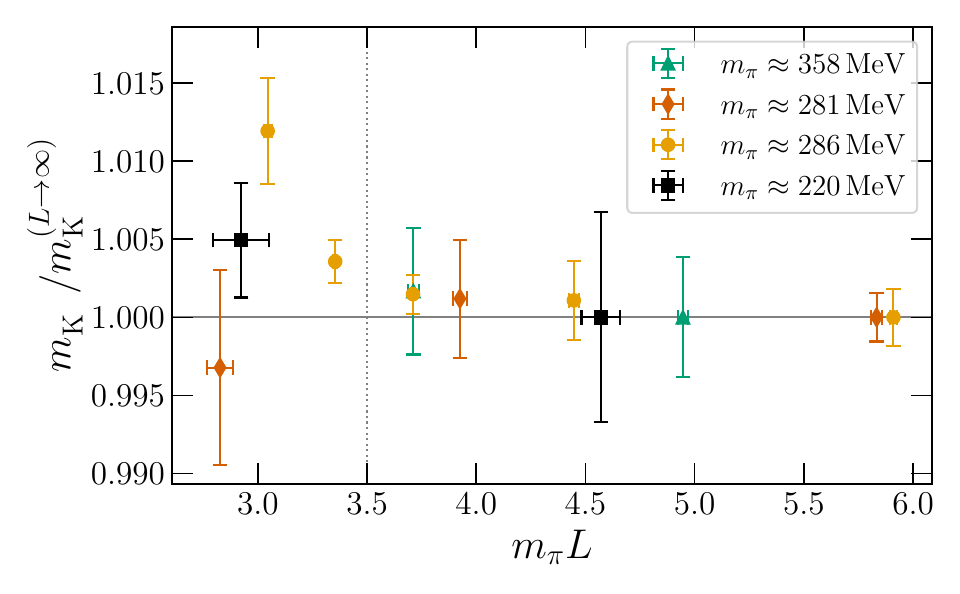}}
	\caption{Volume dependence of the heavy-strange meson mass~(top left) and decay constant~(top right)
        and the pion mass~(bottom left) and kaon mass~(bottom right).
        The results are displayed as in figure \ref{f:fvc}.  }
	\label{f:fvc_extra}
\end{figure}

\newpage 
\section{Scheme dependence \label{a:scheme}}
As outlined in the main text, in particular in \cref{s:scheme}, an ambiguity
with respect to the way isospin symmetric QCD is defined may be present when
comparing results of different collaborations. 
To allow for a comparison of our result with other works
without this ambiguity, we here collect our results based on three
different schemes that differ in the matching of the charm quark mass.

In eq.~(\ref{e:final_results}) we have reported the results using the flavour
averaged D meson mass to fix the mass of the charm quark. 
Using $m_{\rm D_{\rm s}}$ we obtain
\begin{align}
f_{\mathrm{D_s}}=247.0(1.4)\,\mathrm{MeV}\,,\quad f_\mathrm{D}=208.7(1.6)\,\mathrm{MeV}\,,\quad f_{\mathrm{D_s}}/f_\mathrm{D}=1.1833(37)\,,
\end{align}
and with $m_{\eta_{\rm c}}$ we arrive at
\begin{align}
f_{\mathrm{D_s}}=246.9(1.3)\,\mathrm{MeV}\,,\quad f_\mathrm{D}=208.6(1.5)\,\mathrm{MeV}\,,\quad f_{\mathrm{D_s}}/f_\mathrm{D}=1.1838(34)\,.
\end{align}
The impact of switching between the three schemes is thus insignificant.

In table~\ref{t:scaledep} we list the dependence of our final results in the
three different schemes on the input quantities in the form
$S \frac{\partial O}{\partial S}$ for the dependence of observable $O$ on $S$.
The derivatives allow to a posteriori adapt our results to a slightly modified
scheme with respect to the choice used in this work.

\begin{table}[!th]
	\renewcommand*\arraystretch{1.2}
	\centering
	\begin{tabular}{c|*{6}{c}}
		\toprule
		& \multicolumn{6}{c}{$S$} \\ \midrule
		$O$ & $\sqrt{t_0}$ & $m_\pi$ & $m_{\rm K}$ & $m_{\bar{\rm D}}$ & $m_{\rm D_{\rm s}}$ & $m_{\eta_{\rm c}}$
		\\\midrule
		$f_{\rm D_{\rm s}}$ &  $-$1.7329 &  68.9924 & $-$54.5896 &  25.5276 &       -- &       -- \\
		$f_{\rm D}$ &   7.1078 &   4.2036 & $-$65.2799 &  12.8368 &       -- &       -- \\
		$f_{\rm D_{\rm s}}/f_{\rm D}$ & $-$0.048606 & 0.306618 & 0.108713 & 0.049473 &       -- &       -- \\
		
		\hline
		
		$f_{\rm D_{\rm s}}$ &  $-$1.4314 &  65.0836 & $-$55.7642 &       -- &  25.4634 &       -- \\
		$f_{\rm D}$ &   7.1966 &   1.7997 & $-$66.0017 &       -- &  12.7215 &       -- \\
		$f_{\rm D_{\rm s}}/f_{D}$ &  $-$0.047772 & 0.302094 & 0.107553 &       -- & 0.049883 &       -- \\

		\hline
		
		$f_{\rm D_{\rm s}}$ &  $-$1.7556 &  70.6697 & $-$55.3147 &       -- &       -- &  21.3979 \\
		$f_{\rm D}$ &   6.9742 &   5.6292 & $-$65.5603 &       -- &       -- &  10.5432 \\
		$f_{\rm D_{\rm s}}/f_{D}$ &  $-$0.047976 & 0.306910 & 0.106676 &       -- &       -- & 0.042791 \\
		
		\bottomrule

	\end{tabular}
	\caption{Scheme dependence $S \frac{\partial O}{\partial S}$ of observable $O$ with respect to
		the quantity $S$. 
		In the case of $f_{\rm D}$ and $f_{\rm D_{\rm s}}$ the units are MeV, the scheme
		dependence of the ratio is dimensionless.
		The horizontal lines divide three blocks that differ by the matching of the charm quark mass.
		The central values of the six quantities $S$ are
		$\sqrt{t_0} = 0.1449\,$fm,
		$m_\pi = 134.8\,$MeV,
		$m_{\rm K} = 494.2\,$MeV,
		$m_{\bar{\rm D}} = 1899.4\,$MeV,
		$m_{\rm D_{\rm s}} = 1966.0\,$MeV,
		$m_{\eta_{\rm c}} = 2978.0\,$MeV.
		\label{t:scaledep}
	}
\end{table}

\bibliographystyle{jhep}
\bibliography{references}
\end{document}